\documentclass[12pt]{article}

\usepackage{amsmath,amssymb,amsthm}
\usepackage[margin=1in]{geometry}
\usepackage{booktabs}
\usepackage{array}
\usepackage{pgfplotstable}
\pgfplotsset{compat=1.18}
\usepackage{graphicx}
\usepackage{float}
\floatstyle{plaintop}
\restylefloat{figure}
\restylefloat{table}
\makeatletter
\let\lithium@floatc@plain\floatc@plain
\renewcommand\floatc@plain[2]{\begingroup\normalcolor\lithium@floatc@plain{#1}{#2}\endgroup}
\makeatother
\usepackage{placeins}
\usepackage{parskip}
\usepackage{microtype}
\usepackage{xcolor}
\usepackage{enumitem}
\usepackage[round]{natbib}
\usepackage[hidelinks,hyperfootnotes=false]{hyperref}
\makeatletter
\renewcommand{\hyper@natlinkstart}[1]{}
\renewcommand{\hyper@natlinkend}{}
\renewcommand{\hyper@natlinkbreak}[2]{#1}
\makeatother

\usepackage{tikz}
\usetikzlibrary{shapes, arrows.meta, positioning, calc}
\tikzset{link/.style = {draw, -stealth, thick, blue!60!black}}
\definecolor{procomp}{RGB}{34,139,34}
\definecolor{anticomp}{RGB}{178,34,34}
\definecolor{policyblue}{RGB}{31,78,121}

\definecolor{areflink}{rgb}{0.0,0.20,0.55}
\newcommand{\aref}[1]{\hyperref[#1]{\textcolor{areflink}{Appendix~\ref*{#1}}}}
\newcommand{\arefs}[1]{\hyperref[#1]{\textcolor{areflink}{App.~\ref*{#1}}}}

\usepackage{setspace}
\newcommand{\kV}{\kappa^{V}}
\newcommand{\kH}{\kappa^{H}}
\newcommand{\piM}{\pi^{M}} %
\newcommand{\piR}{\pi^{R}} %

\newcommand{\E}{\mathbb{E}}

\newtheorem{lemma}{Lemma}

\newtheorem{assumption}{Assumption}
\newtheorem{proposition}{Proposition}
\newtheorem{definition}{Definition}
\theoremstyle{remark}

\title{Cross-Border Ownership and Input Access:\\
Evidence from the Lithium Supply Chain\thanks{I am grateful to Steven Berry, Pinelopi Goldberg, and Philip Haile for their mentorship. I also thank Charles Hodgson, Amit Khandelwal, Christopher Neilson, and Nicholas Ryan, and participants at the Yale Industrial Organization and Trade Lunch Workshop for helpful comments and discussions. 
I thank Benchmark Mineral Intelligence for providing the data and expertise. This work was supported by the Yale Tobin Center for Economic Policy and the Cowles Foundation.}}
\author{Kei Irizawa\footnote{Department of Economics, Yale University (email: kei.irizawa@yale.edu).}}
\date{November 4, 2026}

\newcommand{\SFtwelveYear}{2024}
\newcommand{\SFtwelveChinaMinedDirectPct}{21}
\newcommand{\SFtwelveChinaOwnedDirectPct}{38}
\newcommand{\SFtwelveChinaShareOfAbroadDirectPct}{22}
\newcommand{\SFtwelveChinaRefinedDirectPct}{72}
\newcommand{\SFtwentythreeYear}{2024}
\newcommand{\SFtwentythreeNMines}{51}
\newcommand{\SFtwentythreeNRefineries}{81}
\newcommand{\SFtwentythreeMineJointPctOneDec}{50.3}
\newcommand{\SFtwentythreeMineJointN}{22}
\newcommand{\SFtwentythreeMineWithinPctOneDec}{69.6}
\newcommand{\SFtwentythreeMineWithinN}{24}
\newcommand{\SFtwentythreeMineAcrossPctOneDec}{56.8}
\newcommand{\SFtwentythreeMineAcrossN}{23}
\newcommand{\SFtwentythreeMineAnyPctOneDec}{85.0}
\newcommand{\SFtwentythreeMineAnyN}{33}
\newcommand{\SFtwentythreeRefJointPctOneDec}{23.0}
\newcommand{\SFtwentythreeRefJointN}{23}
\newcommand{\SFtwentythreeRefWithinPctOneDec}{45.7}
\newcommand{\SFtwentythreeRefWithinN}{37}
\newcommand{\SFtwentythreeRefAcrossPctOneDec}{76.5}
\newcommand{\SFtwentythreeRefAcrossN}{47}
\newcommand{\SFtwentythreeRefAnyPctOneDec}{87.1}
\newcommand{\SFtwentythreeRefAnyN}{58}
\newcommand{\SFtwentyfourYear}{2024}
\newcommand{\SFtwentyfourTopK}{8}
\newcommand{\SFtwentyfourNInBothLists}{5}
\newcommand{\SFtwentyfourTopFourMinePct}{44}
\newcommand{\SFtwentyfourTopFourRefPct}{42}
\newcommand{\SFtwentyfourBothStagesHardRockMinePct}{68}
\newcommand{\SFtwentyfourBothStagesHardRockRefPct}{70}
\newcommand{\SFSourcingYear}{2024}
\newcommand{\SFSourcingOwnerPP}{60.19}

\expandafter\def\csname DTEst@Brine AVC@China\endcsname{2468}
\expandafter\def\csname DTEst@Brine AVC@const\endcsname{3798}
\expandafter\def\csname DTEst@Brine AVC@lagged_evap_anomaly_metres\endcsname{-3142}
\expandafter\def\csname DTEst@Brine AVC@ln_concentration\endcsname{-340}
\expandafter\def\csname DTEst@Brine AVC@ln_electricity\endcsname{621}
\expandafter\def\csname DTEst@Brine AVC@utilization\endcsname{2391}
\expandafter\def\csname DTEst@Demand@linear_time_trend\endcsname{0.075}
\expandafter\def\csname DTEst@Demand@ln_gdp\endcsname{1.911}
\expandafter\def\csname DTEst@Demand@ln_ni\endcsname{0.621}
\expandafter\def\csname DTEst@Demand@ln_oil\endcsname{0.803}
\expandafter\def\csname DTEst@Demand@ln_price\endcsname{-0.511}
\expandafter\def\csname DTEst@Mine AVC@const\endcsname{-27482}
\expandafter\def\csname DTEst@Mine AVC@d_nonspod\endcsname{-1568}
\expandafter\def\csname DTEst@Mine AVC@ln_energy\endcsname{4883}
\expandafter\def\csname DTEst@Mine AVC@ln_grade\endcsname{-3422}
\expandafter\def\csname DTEst@Mine AVC@ln_wage\endcsname{804}
\expandafter\def\csname DTEst@Mine AVC@utilization\endcsname{4947}
\expandafter\def\csname DTEst@Refiner MC@ln_electricity\endcsname{-13}
\expandafter\def\csname DTEst@Refiner MC@ln_wage\endcsname{1607}
\expandafter\def\csname DTEst@Refiner MC@mica\endcsname{982}
\expandafter\def\csname DTEst@Refiner MC@multiplant\endcsname{-4391}
\expandafter\def\csname DTEst@Refiner MC@utilization\endcsname{4847}
\expandafter\def\csname DTSE@Brine AVC@China\endcsname{159}
\expandafter\def\csname DTSE@Brine AVC@const\endcsname{693}
\expandafter\def\csname DTSE@Brine AVC@lagged_evap_anomaly_metres\endcsname{1158}
\expandafter\def\csname DTSE@Brine AVC@ln_concentration\endcsname{160}
\expandafter\def\csname DTSE@Brine AVC@ln_electricity\endcsname{281}
\expandafter\def\csname DTSE@Brine AVC@utilization\endcsname{221}
\expandafter\def\csname DTSE@Demand@linear_time_trend\endcsname{0.037}
\expandafter\def\csname DTSE@Demand@ln_gdp\endcsname{0.526}
\expandafter\def\csname DTSE@Demand@ln_ni\endcsname{0.515}
\expandafter\def\csname DTSE@Demand@ln_oil\endcsname{0.292}
\expandafter\def\csname DTSE@Demand@ln_price\endcsname{0.172}
\expandafter\def\csname DTSE@Mine AVC@const\endcsname{4161}
\expandafter\def\csname DTSE@Mine AVC@d_nonspod\endcsname{614}
\expandafter\def\csname DTSE@Mine AVC@ln_energy\endcsname{552}
\expandafter\def\csname DTSE@Mine AVC@ln_grade\endcsname{434}
\expandafter\def\csname DTSE@Mine AVC@ln_wage\endcsname{234}
\expandafter\def\csname DTSE@Mine AVC@utilization\endcsname{1927}
\expandafter\def\csname DTSE@Refiner MC@ln_electricity\endcsname{3635}
\expandafter\def\csname DTSE@Refiner MC@ln_wage\endcsname{2346}
\expandafter\def\csname DTSE@Refiner MC@mica\endcsname{937}
\expandafter\def\csname DTSE@Refiner MC@multiplant\endcsname{751}
\expandafter\def\csname DTSE@Refiner MC@utilization\endcsname{5495}
\expandafter\def\csname DTScaled@0.1@Refiner MC@utilization\endcsname{485}
\expandafter\def\csname DTScaled@2@Brine AVC@utilization\endcsname{4781}
\expandafter\def\csname DTScaled@2@Mine AVC@utilization\endcsname{9894}
\newcommand{\DTEst}[3][0]{\pgfmathprintnumber[fixed,fixed zerofill,precision=#1,1000 sep={\text{,}}]{\csname DTEst@#2@#3\endcsname}}
\newcommand{\DTSE}[3][0]{\pgfmathprintnumber[fixed,fixed zerofill,precision=#1,1000 sep={\text{,}}]{\csname DTSE@#2@#3\endcsname}}
\newcommand{\DTScaled}[3]{\pgfmathprintnumber[fixed,precision=0,1000 sep={\text{,}}]{\csname DTScaled@#1@#2@#3\endcsname}}
\newcommand{\DTRow}[4][0]{\hspace*{1em}#2 & \DTEst[#1]{#3}{#4} & \DTSE[#1]{#3}{#4} \\}
\newcommand{\CFZeroOneSpotF}{439,749}
\newcommand{\CFZeroOneSpotD}{531,786}
\newcommand{\CFZeroOneSpotChangeAbs}{92,038}
\newcommand{\CFZeroOneSpotPctAbs}{20.93}
\newcommand{\CFZeroOneCaptiveF}{218,555}
\newcommand{\CFZeroOneCaptiveD}{114,005}
\newcommand{\CFZeroOneCaptiveChangeAbs}{104,549}
\newcommand{\CFZeroOneCaptivePctAbs}{47.84}
\newcommand{\CFZeroOneOutputF}{1,074,161}
\newcommand{\CFZeroOneOutputD}{1,074,815}
\newcommand{\CFZeroOneOutputChangeAbs}{654}
\newcommand{\CFZeroOneOutputPctAbs}{0.06}
\newcommand{\CFZeroOneHardrockF}{658,303}
\newcommand{\CFZeroOneHardrockD}{645,792}
\newcommand{\CFZeroOneHardrockChangeAbs}{12,511}
\newcommand{\CFZeroOneHardrockPctAbs}{1.90}
\newcommand{\CFZeroOneBrineF}{415,858}
\newcommand{\CFZeroOneBrineD}{429,023}
\newcommand{\CFZeroOneBrineChangeAbs}{13,165}
\newcommand{\CFZeroOneBrinePctAbs}{3.17}
\newcommand{\CFZeroOnePriceCNF}{12,449.01}
\newcommand{\CFZeroOnePriceCND}{12,434.21}
\newcommand{\CFZeroOnePriceCNChangeAbs}{14.80}
\newcommand{\CFZeroOnePriceCNPctChange}{-0.12}
\newcommand{\CFZeroOneOreF}{7,910.32}
\newcommand{\CFZeroOneOreD}{7,970.41}
\newcommand{\CFZeroOneOreChange}{+60.09}
\newcommand{\CFZeroOneOrePctChange}{+0.76}
\newcommand{\CFZeroOnePricePctAbs}{0.12}
\newcommand{\CFUpdatedTwoBaseOutputPctAbs}{0.06}
\newcommand{\CFUpdatedTwoBaseSpotPctAbs}{20.93}
\newcommand{\CFUpdatedTwoPoolOutputF}{788,526}
\newcommand{\CFUpdatedTwoPoolOutputD}{806,017}
\newcommand{\CFUpdatedTwoPoolOutputChange}{+17,491}
\newcommand{\CFUpdatedTwoPoolOutputPct}{+2.22}
\newcommand{\CFUpdatedTwoPoolOutputPctAbs}{2.22}
\newcommand{\CFUpdatedTwoPoolPriceF}{22,790.08}
\newcommand{\CFUpdatedTwoPoolPriceD}{21,832.73}
\newcommand{\CFUpdatedTwoPoolPriceChange}{-957.35}
\newcommand{\CFUpdatedTwoPoolPricePct}{-4.20}
\newcommand{\CFUpdatedTwoPoolOreF}{4,972.87}
\newcommand{\CFUpdatedTwoPoolOreD}{6,218.30}
\newcommand{\CFUpdatedTwoPoolOreChange}{+1,245.43}
\newcommand{\CFUpdatedTwoPoolOrePct}{+25.04}
\newcommand{\CFUpdatedTwoPoolSpotF}{238,491}
\newcommand{\CFUpdatedTwoPoolSpotD}{231,995}
\newcommand{\CFUpdatedTwoPoolSpotChange}{-6,496}
\newcommand{\CFUpdatedTwoPoolSpotPct}{-2.72}
\newcommand{\CFUpdatedTwoPoolSpotPctAbs}{2.72}
\newcommand{\CFUpdatedTwoPoolCaptiveF}{165,642}
\newcommand{\CFUpdatedTwoPoolCaptiveD}{149,898}
\newcommand{\CFUpdatedTwoPoolCaptiveChange}{-15,744}
\newcommand{\CFUpdatedTwoPoolCaptivePct}{-9.50}
\newcommand{\CFUpdatedTwoPoolHardrockF}{404,133}
\newcommand{\CFUpdatedTwoPoolHardrockD}{381,893}
\newcommand{\CFUpdatedTwoPoolHardrockChange}{-22,240}
\newcommand{\CFUpdatedTwoPoolHardrockPct}{-5.50}
\newcommand{\CFUpdatedTwoPoolBrineF}{384,394}
\newcommand{\CFUpdatedTwoPoolBrineD}{424,124}
\newcommand{\CFUpdatedTwoPoolBrineChange}{+39,731}
\newcommand{\CFUpdatedTwoPoolBrinePct}{+10.34}
\newcommand{\CFUpdatedThreeDivestOutputPctAbs}{0.06}
\newcommand{\CFUpdatedThreeBanOutputPctAbs}{25.36}
\newcommand{\CFUpdatedThreeBanPricePctAbs}{77.19}
\newcommand{\CFUpdatedThreeBanReplacementPct}{41.0}
\newcommand{\CFThreeBanTableSpotF}{439,749}
\newcommand{\CFThreeBanTableSpotB}{133,628}
\newcommand{\CFThreeBanTableSpotChange}{-306,121}
\newcommand{\CFThreeBanTableSpotPct}{-69.61}
\newcommand{\CFThreeBanTableCaptiveF}{218,555}
\newcommand{\CFThreeBanTableCaptiveB}{61,205}
\newcommand{\CFThreeBanTableCaptiveChange}{-157,350}
\newcommand{\CFThreeBanTableCaptivePct}{-72.00}
\newcommand{\CFThreeBanTableOreChinaF}{7,910.32}
\newcommand{\CFThreeBanTableOreChinaB}{20,331.27}
\newcommand{\CFThreeBanTableOreChinaChange}{+12,420.95}
\newcommand{\CFThreeBanTableOreChinaPct}{+157.02}
\newcommand{\CFThreeBanTableOreRoWF}{7,910.32}
\newcommand{\CFThreeBanTableOreRoWB}{2,768.01}
\newcommand{\CFThreeBanTableOreRoWChange}{-5,142.31}
\newcommand{\CFThreeBanTableOreRoWPct}{-65.01}
\newcommand{\CFThreeBanTableOutputF}{1,074,161}
\newcommand{\CFThreeBanTableOutputB}{801,802}
\newcommand{\CFThreeBanTableOutputChange}{-272,360}
\newcommand{\CFThreeBanTableOutputPct}{-25.36}
\newcommand{\CFThreeBanTableHardrockF}{658,303}
\newcommand{\CFThreeBanTableHardrockB}{194,833}
\newcommand{\CFThreeBanTableHardrockChange}{-463,471}
\newcommand{\CFThreeBanTableHardrockPct}{-70.40}
\newcommand{\CFThreeBanTableBrineF}{415,858}
\newcommand{\CFThreeBanTableBrineB}{606,969}
\newcommand{\CFThreeBanTableBrineChange}{+191,111}
\newcommand{\CFThreeBanTableBrinePct}{+45.96}
\newcommand{\CFThreeBanTablePriceF}{12,449.01}
\newcommand{\CFThreeBanTablePriceB}{22,057.83}
\newcommand{\CFThreeBanTablePriceChange}{+9,608.82}
\newcommand{\CFThreeBanTablePricePct}{+77.19}
\newcommand{\CFCurYear}{2024}
\newcommand{\MFERNAll}{139}
\newcommand{\MFERYears}{2016--2024}
\newcommand{\MFERMainRSquared}{0.851}
\newcommand{\MFERMainRMSE}{10.86}
\newcommand{\MFERMainAggregate}{17.25}

\begin{document}
\maketitle
\thispagestyle{empty}

\begin{abstract}
\noindent How does cross-border ownership shape the lithium supply chain? Using global data on mines and refineries, I estimate a model in which ownership can lower refiners' effective ore costs, encourage mines to restrict spot sales, and soften competition. To study these effects, I examine policies motivated by national security concerns about reliable access to critical mineral inputs: divestiture of Chinese stakes in overseas producers and a ban on ore exports to China. Divestiture increases spot ore supply but leaves lithium output nearly unchanged, while continued trade preserves Chinese access to ore. The export ban reduces global lithium supply because refining capacity elsewhere cannot fully replace lost Chinese production. Ownership transfers do not reserve ore for domestic production; restricting Chinese access imposes a substantial supply cost at existing capacities.

\end{abstract}

\setcounter{tocdepth}{2}

\clearpage
\section{Introduction}
\label{sec:intro}

Lithium is a critical mineral used in batteries for electric vehicles and electricity storage. Its importance to these industries has made reliable access to lithium a national security concern. Supply depends on a chain of mining and refining activities spread across countries, with much of refining concentrated in China. Ownership connects these stages across borders: Chinese companies hold interests in mines and refineries at home and abroad. A country's mineral resources therefore do not by themselves determine its refiners' access to ore. Governments have responded to these supply-chain concerns with policies targeting foreign ownership and sourcing relationships. Australia and Canada identify investors' influence over mineral supply chains as a concern in their critical-mineral strategies and investment reviews. The United States also used ownership- and control-based sourcing restrictions in its former clean-vehicle tax credit to limit reliance on designated foreign entities.\footnote{Australian Department of Industry, Science and Resources, \textit{Critical Minerals Strategy 2023--2030}, p.~24; Innovation, Science and Economic Development Canada, ``Policy Regarding Foreign Investments from State-Owned Enterprises in Critical Minerals,'' 28 October 2022; U.S. Department of Energy, ``Final Interpretive Guidance on Foreign Entity of Concern,'' 3 May 2024.}

This paper asks how cross-border ownership shapes access to ore and the supply of lithium. Ownership creates competing profit incentives. A refiner with a stake in a mine shares in the mining profits generated by its ore purchases, which can lower its effective input cost and encourage refining through elimination of double marginalization. A mine whose owners also hold refining interests may restrict sales that would strengthen its affiliates' competitors---an input-foreclosure incentive. Horizontal interests among final lithium producers create incentives to internalize rivals' profits, discouraging expansion that reduces those profits. The effect of cross-border ownership on ore access and lithium supply therefore depends on how these vertical and horizontal interests jointly shape sourcing and production.

To quantify these effects, I develop a model of mining and refining linked by an observed ownership network. Conditional on this network, producers maximize ownership-weighted profits.{\interfootnotelinepenalty=10000\relax\footnote{Section~\mbox{\ref{sec:industry}} presents descriptive evidence motivating the treatment of producers as competitors linked by ownership ties.}} I use policy counterfactuals to examine how ownership incentives shape firms' supply and sourcing decisions. Motivated by governments' concerns about input access, I compare divestiture of Chinese stakes in overseas lithium producers with a ban on ore exports to China. Divestiture changes firms' ownership interests while preserving trade; the export ban restricts ore shipments while retaining ownership.{\interfootnotelinepenalty=10000\relax\footnote{These interventions draw on policy instruments used in practice. Canada ordered Chinese investors to divest stakes in lithium companies (Innovation, Science and Economic Development Canada, statement of 2 November 2022). Zimbabwe suspended lithium-concentrate exports (Ministry of Mines and Mining Development, press statement, February 2026). }} I take these geopolitically motivated interventions as given and study firms' profit-maximizing responses, holding capacity and entry fixed.

The model represents lithium production from hard-rock deposits and brine. Hard-rock mines produce lithium-bearing concentrate, referred to as ore, which refineries purchase and convert into lithium. Brine operations combine extraction and refining to produce lithium from underground salt water without purchasing traded ore. Hard-rock refineries and brine producers compete in the same lithium market. Ownership portfolios connect producers both along the supply chain and across these competing sources of lithium. Measured equity stakes determine the weights each producer places on other mines' and refineries' profits under a maintained proportional-control mapping \citep{OBrienSalop2000, BCS2021}. Hard-rock mines choose how much ore to supply to the spot market, anticipating downstream sourcing and production responses. Hard-rock refiners jointly choose ore purchases and lithium output, sourcing from affiliated mines and the spot market, while brine producers choose lithium output. The ore price clears the spot market, while total refined output determines the lithium price through demand.

I estimate the model's demand and cost parameters using a global panel of mines and refineries for 2015--2024, together with market prices and regional demand. The panel records production, capacity, costs, and ownership. The joint estimation uses regional demand, measured mining costs, and refiners' output optimality conditions. These equations share parameters: mining costs affect refiners' input costs through sourcing, while demand elasticity enters both demand and producers' output decisions. Joint estimation imposes these cross-equation restrictions. Estimation conditions on observed production, while the counterfactuals allow production and sourcing to adjust jointly. I then use the estimated model to answer three questions, solving for sourcing, production, and prices under counterfactual changes in ownership or trade.

I first ask how cross-border ownership shapes access to ore and final lithium supply. To answer this question, I examine divestiture motivated by governments' concerns about reliable access to mineral inputs. Chinese stakes in overseas hard-rock mines and integrated brine operations are transferred to local owners, while trade remains open. Spot ore purchases increase by \CFZeroOneSpotPctAbs\% and the ore price rises, but lithium output increases by only \CFZeroOneOutputPctAbs\%. Divestiture removes affected refiners' cost advantage from sharing in positive mining margins and weakens mines' incentives to restrict spot sales to protect those refiners' profits. Divestiture also weakens the affected brine operation's incentive to restrain output to protect competing refiners' profits. The resulting expansion in brine production largely offsets the contraction in hard-rock refining. Cross-border ownership thus shapes how refiners source ore and which producers supply lithium.

I next ask how downstream competition shapes the effects of cross-border ownership. To answer this question, I compare the same divestiture under observed ownership portfolios and a hypothetical environment in which Chinese owners' refining interests, including their foreign brine holdings, are pooled. This alternative is motivated by the possibility that Chinese industrial policy could encourage coordination to support producer profitability. Relative to each environment's pre-divestiture benchmark, lithium output rises by \CFUpdatedTwoPoolOutputPctAbs\% under pooling, compared with \CFUpdatedTwoBaseOutputPctAbs\% under observed portfolios. Pooling strengthens the affected brine operation's incentive to restrain output to protect a broader set of competing refiners' profits. Divestiture removes the Chinese ownership link carrying these interests, leading to a larger brine expansion. Yet spot ore purchases fall under pooling, reversing their response under observed portfolios. Brine expansion requires no traded ore, while contracting hard-rock refining reduces ore demand. The same ownership intervention can therefore increase or reduce spot trading, depending on how downstream competitive incentives shape production.

Finally, I ask how changing ownership differs from restricting ore trade in its effects on input access and lithium supply. To answer this question, I compare divestiture with a ban on ore exports to China, motivated by governments' concerns about reliable access to domestically mined ore. Divestiture leaves Chinese refiners able to purchase ore through international trade; the ban retains observed ownership but prohibits overseas shipments through both spot and affiliated routes. Both policies are evaluated against a common benchmark with observed ownership and unrestricted ore trade. Divestiture leaves global lithium output nearly unchanged, whereas the ban reduces it by \CFUpdatedThreeBanOutputPctAbs\% and raises the world lithium price by \CFUpdatedThreeBanPricePctAbs\%. Ore prices rise in China and fall outside China. Cheaper ore and a higher lithium price encourage production outside China, but hard-rock refineries there reach capacity. Brine production also expands, yet total production outside China replaces only \CFUpdatedThreeBanReplacementPct\% of China's lost output. The ban therefore makes ore cheaper outside China while reducing global lithium supply: the refineries that retain access cannot expand enough to replace the lost production.

For governments concerned about domestic input access, the counterfactuals show why ownership incentives and refining capacity must be considered together. Divestiture changes both the incentives to supply ore and the demand for it, with responses shaped by competition among final lithium producers. Removing ownership links therefore need not expand spot trading or lower input costs. Continued trade also leaves Chinese refiners able to purchase overseas ore, so transferring ownership does not reserve that ore for domestic production. An export ban prevents overseas ore from reaching Chinese refineries, but replacing the resulting loss of lithium output requires sufficient refining capacity elsewhere. At existing capacities, production outside China can replace only part of the lost Chinese output, so the restriction imposes a substantial cost in global lithium supply.

\subsection*{Related literature}

This paper contributes to the literature on partial ownership and competition. Research on partial and common ownership connects equity holdings to profit internalization among competitors \citep{BS1986, OBrienSalop2000, BCS2021}, while work on vertical integration examines the balance between efficiency gains and foreclosure incentives \mbox{\citep{CLWY2018, HS2025}}. I study a supply chain in which joint ventures and overlapping equity stakes connect producers both across and within production stages. These ownership interests can lower refiners' effective input costs, give mines incentives to restrict spot sales that benefit affiliated refiners' competitors, and discourage expansion by lithium producers that internalize competitors' profits. Because ownership restrictions change these incentives simultaneously, I evaluate them in a joint model of ore supply, sourcing, and lithium production.

The paper also relates to research on resource markets and battery supply chains. \citet{ACDL2019} quantify how market power distorts the allocation of global oil extraction. \citet{DEH2026} quantify how resource concentration and the joint use of minerals in battery production shape policy spillovers across critical-mineral markets. \mbox{\citet{BKLZ2025}} study how learning in battery production shapes the effects of electric-vehicle subsidies and local-content requirements. I complement this work by quantifying how cross-border ownership shapes competition and the supply of lithium, a key input into battery production.

Finally, the paper relates to research on geoeconomics and the use of national policies to reshape international production and trade. \citet{CMS2026} study how governments use access to economic relationships to exercise international influence. \mbox{\citet{BKZ2025}} quantify how different industrial-policy instruments affect firms' decisions and global industry outcomes. I examine geopolitically motivated restrictions on foreign ownership and ore trade. Divestiture changes producers' profit incentives while leaving trade open; an ore-export ban restricts shipments while leaving ownership unchanged. Comparing these interventions allows me to quantify how vertical and horizontal ownership interests shape firms' responses to policies targeting cross-border relationships.

Section~\ref{sec:industry} describes the lithium industry, ownership relationships, and policy motivations. Section~\ref{sec:data} presents the data. Section~\ref{sec:model} develops the model, and Section~\ref{sec:identification} explains identification of model parameters. Section~\ref{sec:results} reports the parameter estimates and assesses model fit. Section~\ref{sec:counterfactuals} examines divestiture, the role of downstream competition, and the comparison with an ore-export ban. Section~\ref{sec:conclusion} concludes.

\section{Industry structure and ownership}
\label{sec:industry}\label{sec:institutions}\label{sec:industry:valuechain}

The geography of lithium production and cross-border ownership help explain governments' concerns about mineral access. The effects of ownership and trade restrictions on lithium supply depend on how mines adjust ore supply and how refiners adjust sourcing and production. Ownership links shape these responses by giving producers incentives to account for the profits of firms at other production stages and of their competitors. This section describes the industry and policy setting, documents these ownership relationships, and examines their association with sourcing.

The paper studies mining and refining, which supply lithium chemicals for downstream cathode and battery production. Two production routes connect these stages. Hard-rock mines produce a concentrate, referred to as \textit{ore}, that can be traded and shipped to refineries for conversion into lithium chemicals, referred to as \textit{lithium}. Brine operations combine extraction and refining to produce lithium from underground salt water, without a separate traded-ore sourcing decision. Hard-rock refiners and brine producers compete in the same lithium market, so changes in ore availability can affect brine producers through their effects on hard-rock output and lithium prices.

Production sites are distinct from their owners. An owner may hold interests in several mines and refineries, including sites in different countries and using different production routes. Understanding mineral access therefore requires examining both where production takes place and the ownership relationships connecting producers.

\subsection{Production geography and mineral-access policy}
\label{sec:industry:geography}\label{sec:policy_object}
\label{sec:industry:facts}\label{sec:data_facts}

\begin{figure}[!htbp]
\centering
\includegraphics[width=\textwidth]{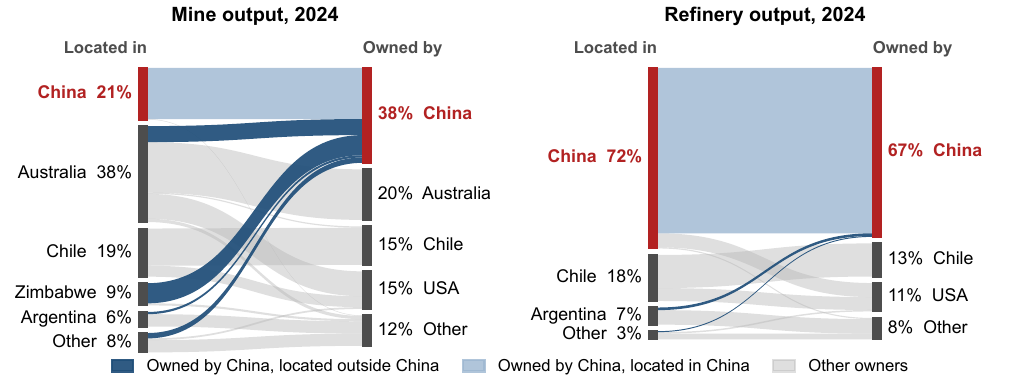}
\caption{Lithium production by location and owners' nationality (\SFtwelveYear).}
\label{fig:geography}
\begin{minipage}{0.95\textwidth}
\vspace{0.3em}
\footnotesize \textit{Notes:} The left side of each panel shows where production takes place; the right side attributes that output to corporate owners' nationalities using their equity shares in each producing site. Band widths measure shares of global output at the corresponding production stage. Shares use LCE production.
Dark blue bands show production outside China attributed to Chinese owners; light blue bands show production in China attributed to Chinese owners. 
Output attributed to a corporate owner is not further allocated to its non-controlling shareholders. 
Smaller country shares and unattributed ownership are grouped into ``Other.''
\end{minipage}
\end{figure}

Mining and refining have different geographical distributions. In \SFtwelveYear{}, China accounts for \SFtwelveChinaRefinedDirectPct\% of refinery output and \SFtwelveChinaMinedDirectPct\% of mine output, while Australia and Chile supply much of the mined lithium. The left side of each panel in Figure~\ref{fig:geography} describes the location of production at that stage. The right side attributes the same production to owners' nationalities in proportion to equity stakes. On this basis, Chinese owners account for \SFtwelveChinaOwnedDirectPct\% of world mine output. Their overseas stakes represent \SFtwelveChinaShareOfAbroadDirectPct\% of output mined outside China, highlighted by the dark bands. China's position in the supply chain thus combines refining within China with ownership interests in mining abroad.

These geographical and ownership patterns help explain why governments scrutinize ownership when considering mineral access. A mine located within a country can be connected to the sourcing needs and profit incentives of refiners abroad. Domestic extraction therefore need not ensure access for domestic refiners. Concentrated refining capacity adds another source of dependence: ore must reach facilities able to convert it into lithium chemicals.

Governments have responded to these mineral-access concerns with policies targeting ownership and sourcing. Australia's Critical Minerals Strategy warns that foreign ownership and offtake arrangements could limit Australian refiners' access to Australian minerals. Investment screening can restrict the acquisition of such interests: Australia prohibited a China-linked investor from taking full ownership of the company owning the Bald Hill lithium mine.\footnote{Australian Department of Industry, Science and Resources, \textit{Critical Minerals Strategy 2023--2030}, p.~24; \textit{Foreign Acquisitions and Takeovers (Prohibition of Proposed Action) Order (No.~2) 2023}, 20 July 2023. } Canada's investment policy considers an investor's influence over the supply chain, alternative sources of supply, and where minerals will be processed or exported. It has also required Chinese investors to divest stakes in lithium companies following national-security reviews.\footnote{Innovation, Science and Economic Development Canada, ``Policy Regarding Foreign Investments from State-Owned Enterprises in Critical Minerals under the Investment Canada Act,'' 28 October 2022; statement of the Minister of Innovation, Science and Industry, 2 November 2022. } The United States used sourcing eligibility to encourage domestic and partner-country production and reduce reliance on China and other designated countries. Its former clean-vehicle tax credit made vehicles with batteries containing critical minerals extracted, processed, or recycled by foreign entities of concern ineligible for the credit, with that classification based partly on ownership and control.\footnote{U.S. Department of Energy, ``Interpretation of Foreign Entity of Concern,'' 89 Fed.\ Reg.\ 37079, 37082 (6 May 2024); 26 U.S.C.\ \S~30D(d)(7). The credit ended in 2025 under Public Law 119--21.}

\begingroup
\clubpenalty=10000
Governments have also intervened directly in ore trade. Zimbabwe has restricted lithium-concentrate exports to encourage domestic processing. Given China's dominant role in refining and Chinese firms' substantial ownership interests in Zimbabwean mines, these measures also affect the supply links connecting overseas mines to Chinese refiners.\footnote{Zimbabwe suspended lithium-concentrate exports in February 2026, then allowed approved producers to resume shipments under quotas in April. See Ministry of Mines and Mining Development, “Press Statement on Ban of Export of Raw Minerals and Lithium Concentrate,” February 2026; Reuters, “China's Yahua Industrial Group gets lithium export quota from Zimbabwe,” 14 April 2026.}
\par\endgroup

Security concerns may be forward-looking: governments may seek to limit exposure to future disruption or coercion even when firms currently maximize profits. The analysis takes the interventions as given and evaluates firms' responses at existing capacities. The paper compares divestiture of Chinese stakes in overseas mines and integrated brine operations with an ore-shipment restriction to examine their consequences for sourcing, production, and prices. These are stylized interventions motivated by the policy concerns above, rather than replications of any individual policy. To understand firms' responses to these interventions, I next describe the ownership relationships linking mines to refineries and connecting competing producers.

\FloatBarrier

\subsection{Joint ventures and overlapping ownership}
\label{sec:industry:portfolios}

Joint ventures and shared ownership create links among producers. A joint venture has multiple owners. Horizontal links arise when an owner holds stakes in competing mines or competing refineries; vertical links connect mines and refineries through a shared owner. The paper describes a site as vertically integrated when it has such a link to the other stage, including through partial ownership, and vertically independent otherwise. These ownership structures can overlap: a jointly owned mine can also share owners with other mines and with refineries.

Table~\ref{tab:ownership_config} shows how prevalent these ownership structures are among producing sites. Sites that are jointly owned or have horizontal or vertical ownership links account for \SFtwentythreeMineAnyPctOneDec\% of mining output and \SFtwentythreeRefAnyPctOneDec\% of refinery output. The link measures exclude pairs of sites with identical ownership vectors, so they capture relationships beyond wholly shared portfolios. The overlapping categories show why vertical ownership links must be considered alongside horizontal ownership links among competitors.

\begin{table}[!htb]
\centering
\small
\caption{Joint ventures and ownership links among producing sites, \SFtwentythreeYear.}
\label{tab:ownership_config}
\begin{tabular}{@{}lrrrr@{}}
\toprule
 & \multicolumn{2}{c}{Mines} & \multicolumn{2}{c}{Refineries} \\
\cmidrule(lr){2-3}\cmidrule(l){4-5}
Ownership structure & Output (\%) & Sites & Output (\%) & Sites \\
\midrule
Joint venture & \SFtwentythreeMineJointPctOneDec & \SFtwentythreeMineJointN & \SFtwentythreeRefJointPctOneDec & \SFtwentythreeRefJointN \\
Horizontal ownership link & \SFtwentythreeMineWithinPctOneDec & \SFtwentythreeMineWithinN & \SFtwentythreeRefWithinPctOneDec & \SFtwentythreeRefWithinN \\
Vertical ownership link & \SFtwentythreeMineAcrossPctOneDec & \SFtwentythreeMineAcrossN & \SFtwentythreeRefAcrossPctOneDec & \SFtwentythreeRefAcrossN \\
Any of the above & \SFtwentythreeMineAnyPctOneDec & \SFtwentythreeMineAnyN & \SFtwentythreeRefAnyPctOneDec & \SFtwentythreeRefAnyN \\
\midrule
All producing sites & & \SFtwentythreeNMines & & \SFtwentythreeNRefineries \\
\bottomrule
\end{tabular}
\begin{minipage}{0.95\textwidth}
\vspace{0.4em}
\footnotesize \textit{Notes:} The table covers hard-rock and brine mines and refineries.
A joint venture has more than one owner. Horizontal links connect sites at the same production stage through a shared owner; vertical links connect mines and refineries through a shared owner. Both link measures exclude pairs with exactly the same owners and equity shares, including a brine operation's own mine--refinery pair. Each output share counts the full 
LCE production of sites in that category.
The categories are not mutually exclusive: a site may have multiple ownership links and appear in more than one row. The reported shares are therefore not additive.
\end{minipage}
\end{table}

These ownership structures are widespread, and several major owners hold substantial interests at both stages. Table~\ref{tab:top_owners} identifies these owners and reports their equity-attributed shares of mining and refining output. In \SFtwentyfourYear{}, the four largest owners accounted for \SFtwentyfourTopFourMinePct\% of mining output and \SFtwentyfourTopFourRefPct\% of refinery output when each site's production is allocated by equity share; \SFtwentyfourNInBothLists{} of the \SFtwentyfourTopK{} largest owners at each stage appear in both lists. Brine contributes mechanically to this overlap because the same integrated producer spans mining and refining. The pattern also extends to hard rock: owners with interests at both stages hold \SFtwentyfourBothStagesHardRockMinePct\% of hard-rock mining output and \SFtwentyfourBothStagesHardRockRefPct\% of hard-rock refining output.

\begin{table}[!htb]
\centering
\small
\caption{Major owners' shares of mining and refining output, \SFtwentyfourYear.}
\label{tab:top_owners}
\begin{tabular}{@{}lrr@{}}
\toprule
Corporate owner & Mine output (\%) & Refinery output (\%) \\
\midrule
SQM & 15.2 & 13.2 \\
Albemarle & 14.6 & 11.2 \\
\textcolor{anticomp}{Ganfeng} & 3.8 & 11.0 \\
\textcolor{anticomp}{Tianqi} & 4.1 & 5.4 \\
Arcadium Lithium & 4.5 & 3.8 \\
Pilbara Minerals & 7.9 & 0.3 \\
\textcolor{anticomp}{Canmax} & -- & 6.6 \\
\textcolor{anticomp}{Chengxin} & 2.1 & 4.2 \\
\textcolor{anticomp}{Sinomine} & 2.9 & 3.2 \\
Mineral Resources & 5.8 & -- \\
\bottomrule
\end{tabular}
\begin{minipage}{0.95\textwidth}
\vspace{0.4em}
\footnotesize \textit{Notes:} Each owner's attributed output is the sum of production at each site multiplied by the owner's equity share in that site. The reported percentages divide this attributed output by global mining or refining output, respectively, including both hard-rock and brine production. Shares use reported LCE production.
Output attributed to a corporate owner is not further allocated to its non-controlling shareholders. 
Rows are ordered by the sum of the mining and refining output shares. Red names identify Chinese owners. A dash denotes no attributed output at that stage.
\end{minipage}
\end{table}

Portfolios also span production routes. Ganfeng and Tianqi hold interests in hard-rock mines, hard-rock refineries, and brine operations.\footnote{Ganfeng Lithium, \textit{2018 Annual Report}; Tianqi Lithium, \textit{2023 Annual Report}.} For a hard-rock refiner, a stake in a hard-rock mine creates a financial interest in an ore supplier, while a stake in a brine operation creates an interest in a competing lithium producer. The project portfolios retained in the model combine vertical links to input suppliers with horizontal links to competitors in the lithium market.

These ownership links connect Chinese producers that also compete with one another. Tianqi's 2022 prospectus describes competition with domestic and international producers and identifies Ganfeng among the major suppliers in its discussion of competition.\footnote{Tianqi Lithium Corporation, \textit{2022 Prospectus}%
.} These disclosures motivate treating Chinese producers as firms with distinct profit incentives, while allowing ownership links to shape how they value other producers' profits. Section~\mbox{\ref{sub:cf-q2}} considers a counterfactual in which Chinese owners' refining interests are pooled, allowing producers to internalize a broader set of competitors' profits.

Upstream ownership can strengthen the competitive position of affiliated refiners. Tianqi describes its interest in Greenbushes as providing a stable supply of low-cost lithium concentrate for its refining operations. It presents resource access and vertical integration as competitive strengths and attributes improvements in cost efficiency and profitability to integration across mining and processing.\footnote{Tianqi Lithium Corporation, \textit{2022 Prospectus}%
.} This account links upstream ownership to reliable access to low-cost ore and motivates examining how ownership affects refiners' sourcing and production decisions.

\begin{samepage}
The same ownership links can also influence whether a mine supplies unaffiliated refiners. In 2024, Greenbushes was jointly owned by Albemarle and a joint venture between Tianqi and IGO, linking the mine to their lithium refineries. On IGO's quarterly results call, its chief executive Ivan Vella explained the decision about sales to other buyers:
\begin{quote}
``And at this point, we've taken a decision to not proceed with any third-party sales. The market is still very uncertain at this point, and we're obviously looking at the business and optimizing the production, but also making sure that we maintain strength in that overall lithium market as we think about it in an integrated manner.''\footnote{Ivan Vella, Chief Executive Officer, IGO Limited, Q2 FY2024 earnings call, 31 January 2024, p.~9.}
\end{quote}
\end{samepage}

The statement describes production and sales decisions that take the integrated business into account. Selling ore to an unaffiliated refiner can generate mining profits while strengthening a competitor of the mine's affiliated refineries, creating a potential incentive to limit those sales. The quotation motivates allowing mine decisions to reflect refining interests. Horizontal ownership links matter in turn because the wider portfolio determines how owners value downstream gains and losses arising from additional ore supply.

The next subsection examines how these ownership relationships are reflected in refiners' access to ore.

\subsection{Ownership and sourcing relationships}
\label{sec:industry:sourcing}

As descriptive evidence, I examine whether mines and refineries that share an owner are more likely to have a reported sourcing relationship, controlling for geography and whether both producers have Chinese owners. I construct potential mine--refinery pairs from the \SFSourcingYear{} production and ownership data and identify which pairs have a reported supplier relationship. I estimate linear probability models. The most extensive specification is
\begin{equation}
L_{mr}=\alpha_m+\gamma_r+\lambda_{c(m),c(r)}+\beta S_{mr}+\delta C_m C_r+\theta\log d_{mr}+\varepsilon_{mr}.
\label{eq:ownership_sourcing}
\end{equation}
Here, $L_{mr}$ indicates whether mine $m$ supplies refinery $r$, $S_{mr}$ indicates a shared identified owner, and $C_m$ and $C_r$ indicate a positive identified Chinese ownership stake. The remaining terms are mine fixed effects $\alpha_m$, refinery fixed effects $\gamma_r$, fixed effects for the ordered pair of host countries $\lambda_{c(m),c(r)}$, and log distance $\log d_{mr}$. 

\begin{table}[!t]
\centering
\small
\caption{Ownership and reported hard-rock supply relationships.}
\label{tab:ownership_sourcing}
\begin{tabular}{lccc}
\toprule
 & (1) & (2) & (3) \\
\midrule
Shared owner & $0.603$ & $0.600$ & $0.602$ \\
 & $(0.088)$ & $(0.089)$ & $(0.089)$ \\
Chinese ownership on both sides &  &  & $-0.031$ \\
 &  &  & $(0.018)$ \\
Log distance (km) & $-0.011$ & $-0.019$ & $-0.020$ \\
 & $(0.005)$ & $(0.007)$ & $(0.007)$ \\
\midrule
Mine fixed effects & Yes & Yes & Yes \\
Refinery fixed effects & Yes & Yes & Yes \\
Host-country-pair fixed effects & No & Yes & Yes \\
Mines & 33 & 33 & 33 \\
Refineries & 45 & 45 & 45 \\
Pairs & 1{,}485 & 1{,}485 & 1{,}485 \\
Reported links & 55 & 55 & 55 \\
\bottomrule
\end{tabular}
\begin{minipage}{0.95\textwidth}
\vspace{0.4em}
\footnotesize \textit{Notes:} The sample pairs all producing hard-rock mines with producing hard-rock refineries for which complete supplier lists are available, using the \SFSourcingYear{} production and ownership panel. Each observation is a potential mine--refinery pair. The outcome equals one if the mine supplies the refinery and zero otherwise. 
``Shared owner'' equals one if the same owner holds a positive stake in both sites. ``Chinese ownership on both sides'' equals one if each site has a positive stake held by an owner from China; the owners need not be the same. 
\end{minipage}
\end{table}

Table~\ref{tab:ownership_sourcing} shows that shared ownership is strongly associated with sourcing. The estimated relationship remains similar across specifications, including those accounting for host-country pairs and Chinese ownership on both sides. This stability suggests that the ownership--sourcing association extends beyond the geographical concentration of mining and refining. In the most extensive specification, sharing an owner is associated with a \SFSourcingOwnerPP{} percentage-point higher probability of a supply relationship.

Chinese ownership on both sides does not add a positive association conditional on these factors. This finding suggests that sourcing relationships are associated with shared ownership beyond any general preference for sourcing between producers with Chinese owners.
Yahua's experience at Greenbushes illustrates this distinction. In a 2018 regulatory response, Yahua reports that its refining subsidiary, Xingsheng Lithium, had previously sourced most of its lithium concentrate from Greenbushes. As the mine prioritized supply to its owners, Tianqi and Albemarle, and largely ceased external sales, Xingsheng's concentrate purchases fell in 2015, and it was unable to purchase any concentrate throughout 2016.\footnote{Yahua Group, \textit{Revised Response to Feedback on the Convertible Bond Application}, 16 November 2018%
. The filing refers to Greenbushes' operator, Talison.} Chinese ownership of the mine therefore did not ensure access for this unaffiliated Chinese refiner.

Together, the company accounts and descriptive sourcing patterns motivate the model developed below, in which ownership links shape refiners' access to ore. The following sections describe the data (Section~\ref{sec:data}) and develop the model (Section~\ref{sec:model}) used to quantify the effects of ownership changes on sourcing, production, and prices.

\FloatBarrier
\section{Data}
\label{sec:data}

I assemble a panel of lithium mines and refineries covering global production. The panel records production, capacity, costs, and ownership, which I combine with market prices and regional demand. This section describes the data sources, sample coverage, and measurement of the main variables. Section~\ref{sec:identification} explains how these data enter demand and cost estimation.

\subsection{Sources, sample, and coverage}
\label{sec:data_sources}

The panel spans 2015--2024 and draws primarily on Benchmark Mineral Intelligence (BMI) and S\&P Capital IQ Pro (S\&P), supplemented by company filings and public sources. The unit of observation is the operator-site: a mine or refinery aggregated to the operating entity at a single location. The panel distinguishes hard-rock mines, hard-rock refineries, and integrated brine producers. Brine extraction and refining are co-located, so the summary statistics report production, capacity, and utilization once for each brine producer's site. Reported production and capacity for both ore and lithium are expressed in tonnes of lithium carbonate equivalent (LCE), which converts each product's lithium content into an equivalent quantity of lithium carbonate. Descriptive statistics use these reported quantities and capacities. For the model, I multiply hard-rock mine quantities and capacities by a common reference recovery rate to express them in recoverable-LCE units, leaving refinery quantities and capacities unchanged. Ore prices and mining costs use this recoverable-LCE basis.

\subsection{Production, capacity, and costs}
\label{sec:data_sumstats}

The operating records contain annual production, capacity, feedstock technology, and location. Nameplate capacity measures annual output under normal operating conditions. BMI's average variable costs cover production through saleable ore for hard-rock mines and through lithium chloride for brine operations. Subsequent conversion into refined lithium is accounted for separately.

I supplement these records with wages, electricity prices, geological characteristics, and meteorological conditions. Site locations, established from regulatory records and company documents, link producers to regional factor prices and brine sites to local weather data. Hard-rock head grades come from BMI and measure the lithium content of mined rock. Source-brine concentrations measure lithium per unit of fluid and are assembled from technical filings, geological studies, and company disclosures. Wages measure sectoral earnings, using NBS/CEIC provincial series for China and ILOSTAT and national sources elsewhere. Electricity prices come from official regional and national industrial-price and tariff series.

\subsection{Ownership}
\label{sec:data:sourcing}

I use BMI and S\&P as the primary sources for annual ownership stakes, supplemented by corporate filings, transaction records, and historical shareholder records. These records identify the owners of each operator-site and their equity shares.

I standardize owner names across assets and group each parent company with the subsidiaries it controls. I trace project holding vehicles to their corporate owners and retain each owner's equity stake in jointly owned operations. I do not further allocate a corporate owner's project portfolio to its non-controlling shareholders. Project-level minority stakes therefore enter the ownership network, whereas non-controlling shareholdings between corporate owners do not generate additional indirect project interests. State-affiliated entities are recorded separately, so shared state affiliation alone does not create an ownership link. The ownership links studied here arise from operating companies' stakes in suppliers and competitors, whereas \citet{ASTU2018,BCS2021} emphasize overlapping institutional-investor holdings. 

As supplementary evidence, I use BMI data on mine--refinery supply relationships in 2024. These data identify which mines supply each refinery but do not report shipment quantities or transaction prices. 

\subsection{Prices and demand}
\label{sec:data:markets}

The market series comprise an ore price benchmark (FOB Australia), a lithium carbonate price benchmark (CIF Asia), and regional lithium demand. Demand is measured at the region-year level in LCE. BMI constructs it from final consumption of battery and non-battery products: battery demand is converted into lithium content using the cathode-chemistry mix and the lithium requirements of each chemistry, and non-battery uses are added. The inputs reflect realized end-use activity over the sample period. The regions are Africa, China, Europe, Asia excluding China, the Middle East, Oceania, North America, and South and Central America.

\FloatBarrier

\section{Model}
\label{sec:model}

The model maps owners' equity portfolios into the objectives of ore and lithium producers. Under proportional control, each producer maximizes a weighted sum of its owners' portfolio profits, with weights proportional to their equity stakes in that producer. Ownership links between mines and refiners lead producers to internalize profits across the two stages; ownership links among competing producers lead them to internalize rivals' profits.

Profit maximization is a maintained assumption about operating decisions conditional on ownership; the ownership structure may itself reflect government intervention.

Hard-rock mines choose spot commitments, anticipating how refiners' responses affect their own and affiliated profits. Hard-rock refiners choose ore sources and lithium output, while brine producers compete in the same lithium market without buying traded ore. Vertical profit interests can give mines an incentive to restrict spot supply and refiners an incentive to expand production through reduced double marginalization. Purchasing limits constrain the quantities refiners can obtain under ownership-linked rights.

I first explain how ownership determines producers' profit objectives and describe the ore-purchasing arrangements and equilibrium conditions. I then characterize downstream sourcing and production decisions before examining upstream incentives to supply ore.

\subsection{Ownership and producer objectives}
\label{sec:environment}
\label{sec:ownership_map}

\subsubsection{Producers, ownership, and demand}
\label{sec:owners_producers}
Owners $c\in\mathcal C$ hold equity stakes $\omega_{cf}$ in producers $f$, comprising hard-rock mines, hard-rock refiners, and brine producers. An owner may hold stakes in several producers, and a producer may have several owners.

A hard-rock mine $m\in\mathcal M$ has total ore output $x_m$ up to capacity $K_m$, with variable cost $C^M_m(x_m)$ and marginal mining cost $C_m^{M\prime}(x_m)$. It chooses the quantity committed to the spot market, which clears at price $w$. A hard-rock refiner $r\in\mathcal R$ purchases ore and chooses lithium output $q_r$ up to capacity $K_r$, selling it at price $P$. Lithium output is measured in tonnes of lithium carbonate equivalent (LCE). With ore measured in recoverable-LCE units under the common recovery assumption, producing \mbox{$q_r$} units of lithium requires \mbox{$q_r$} units of ore.

Brine co-locates extraction and refining. Its intermediate is typically not traded, so a brine producer takes no separate decision in the ore market: it neither buys ore nor commits ore to the spot market. It enters $\mathcal R$ as a downstream supplier of lithium with a bundled extraction and refining cost. Thus hard-rock mines choose spot ore supply, while hard-rock refiners and brine producers compete in the lithium market.

\phantomsection\label{subsec:company_agg}
The index $r$ identifies an individual downstream producer, either a hard-rock refiner or a brine producer. Write $\mathcal R^{\mathrm{HR}}$ and $\mathcal R^{\mathrm{BR}}$ for the two subsets of $\mathcal R$. Downstream producers with identical ownership vectors form a single strategic agent, which chooses their outputs and ore purchases jointly. Each producer retains its own technology, costs, capacity, and purchasing constraints. I refer to these producers as sister plants; producers with different ownership vectors remain separate strategic agents.

Refined lithium is a homogeneous product traded in a global market at a common price $P_t$. Let $Q_t=\sum_{r\in\mathcal R} q_{rt}$ denote global output in year $t$ across hard-rock refiners and brine producers. {Demand in this market aggregates the regional consumption demands specified in Section~\mbox{\ref{est-sec:demand}}. } Global inverse demand is
\begin{equation}
  P_t(Q_t)=\left(\frac{Q_t}{A_t}\right)^{1/\eta}
  \label{eq:inverse_demand}
\end{equation}
where $A_t$ determines the level of demand and $\eta$ is the price elasticity of demand. Write
\[
  b_t(Q_t)\equiv-P_t'(Q_t)=\frac{P_t(Q_t)}{|\eta|Q_t}
\]
for the magnitude of the inverse-demand slope. 

\subsubsection{From owner portfolios to producer objectives}
\label{sec:producer_objectives}
Let $\piM_m$ and $\piR_r$ denote the operating payoffs of mine $m$ and downstream producer $r$, specified in the producer problems below. Owner $c$ values its portfolio payoff
\begin{equation}
  V_c=\sum_m\omega_{cm}\piM_m+\sum_r\omega_{cr}\piR_r.
  \label{eq:owner_payoff}
\end{equation}
I adopt the proportional-control benchmark of the partial- and common-ownership literature \citep{OBrienSalop2000,BCS2021}.
\begin{assumption}[Proportional control]\label{ass:proportional_control}
Producer $f$ weights each owner's portfolio payoff by that owner's equity stake in the producer, maximizing $\sum_c\omega_{cf}V_c$.
\end{assumption}

For downstream producer $r$, substituting the owners' portfolio payoffs gives
\begin{equation}
  \sum_c\omega_{cr}V_c
  =\left(\sum_c\omega_{cr}^2\right)\piR_r
  +\sum_{j\neq r}\left(\sum_c\omega_{cr}\omega_{cj}\right)\piR_j
  +\sum_m\left(\sum_c\omega_{cr}\omega_{cm}\right)\piM_m.
  \label{eq:Vr_unnorm}
\end{equation}
Dividing the objective by the positive coefficient on own profit, $\sum_c\omega_{cr}^2$, normalizes the weight on $\piR_r$ to one without changing the maximizer. The same calculation for any producer $f$ gives its weight on producer $g$:
\begin{equation}
  \kappa_{fg}=\frac{\sum_c\omega_{cf}\omega_{cg}}
                       {\sum_c\omega_{cf}^{2}}.
  \label{eq:kappa_def}
\end{equation}
I call $\kappa_{fg}$ a profit-internalization weight and write $\kV_{fg}$ for links across stages and $\kH_{fg}$ for links within a stage. The strategic objectives are
\begin{align}
  V_r
  &=\piR_r+\sum_{j\neq r}\kH_{rj}\piR_j+\sum_m\kV_{rm}\piM_m,
  \label{eq:Vr}\\
  V_m
  &=\piM_m+\sum_{k\neq m}\kH_{mk}\piM_k+\sum_r\kV_{mr}\piR_r.
  \label{eq:Vm}
\end{align}
The first subscript of each weight identifies the producer making the decision. Weights are directional, need not be symmetric, and can exceed one; they are profit-internalization weights, whereas equity shares allocate operating payoffs as in~\eqref{eq:owner_payoff}. The weight $\kappa_{fg}$ measures how much producer $f$ values an increase in producer $g$'s profit relative to the same increase in its own profit. The vertical weight $\kV_{rm}$ therefore makes refiner $r$ account for the mining profits generated by its purchases from mine $m$, while $\kV_{mr}$ makes mine $m$ account for how its supply affects downstream producer $r$'s profits. The horizontal weights $\kH_{rj}$ and $\kH_{mk}$ similarly make downstream refiners and upstream mines account for their effects on competing producers' profits.

In disclosed lithium joint ventures, shareholder representation is linked to equity, motivating the use of stakes to summarize influence.\footnote{Under the Wodgina terms announced in 2019, management-committee representation was proportional to the partners' interests, while certain decisions required both parties' approval and matters involving conflicts of interest were assigned to the non-conflicted partner's representatives (Albemarle, Form 8-K, 1~August 2019, Item~1.01). At Minera Exar, the announced move to equal ownership on closing included equal representation on the board and management committee (Lithium Americas, press release, 1~April 2019, ``Transaction Details'').} At Greenbushes, the shareholder-appointed board approves business plans, budgets, and capital expenditure, and production is planned with the shareholders' needs in view.\footnote{IGO Limited, \textit{ASX JV Presentation}, 9~December 2020; Tianqi Lithium, FY2020 earnings call, 7~May 2021.} These arrangements do not establish exact proportionality between equity and influence: voting thresholds, minority protections, and conflict rules can separate the two. Proportional control is therefore an approximation for aggregating owners' interests.

Joint-venture partners share an interest in the producer's profits, but their holdings in other mines and refiners can lead them to favor different production choices. For a jointly owned refiner, for example, one partner may hold a stake in a supplying mine, while another may own a competing refiner. The partners therefore account for different effects of the refiner's production on profits elsewhere in their portfolios. Under proportional control, the partners' equity stakes determine how these different portfolio interests enter the refiner's objective.

\begin{figure}[!htbp]
\centering
\begin{minipage}{0.46\textwidth}
\centering
\textbf{(a) Vertical ($\kV$)}\\[0.4em]
\begin{tikzpicture}[scale=1.25, transform shape]
  \node[draw, rounded corners, thick, minimum height=0.8cm, minimum width=1.5cm,
        text centered, font=\footnotesize,
        path picture={
          \fill[anticomp!30] (path picture bounding box.south west) rectangle ([xshift=0.75cm]path picture bounding box.south west |- path picture bounding box.north);
          \fill[policyblue!30] ([xshift=0.75cm]path picture bounding box.south west) rectangle (path picture bounding box.north east);
        }] (rA) {Refiner $r$};
  \node[draw, rounded corners, thick, minimum height=0.8cm, minimum width=1.5cm,
        text centered, font=\footnotesize, right=2.2cm of rA, fill=policyblue!30] (rB) {Refiner $j$};
  \node[draw, circle, thick, minimum size=0.9cm, text centered, font=\footnotesize,
        fill=anticomp!30, below=1.8cm of $(rA)!0.5!(rB)$] (m1) {Mine $m$};
  \node[font=\scriptsize, above=0.1cm of rA] {A: 50\% \; B: 50\%};
  \node[font=\scriptsize, above=0.1cm of rB] {B: 100\%};
  \node[font=\scriptsize, below=0.1cm of m1] {A: 100\%};
  \draw[-stealth, thick, blue!60!black] (rA) to[bend left=20]
      node[midway, right=2pt, font=\scriptsize, text=blue!60!black,
           fill=white, inner sep=1pt] {$\kV_{rm}{=}1$} (m1);
  \draw[-stealth, thick, blue!35!black] (m1) to[bend left=20]
      node[midway, left=2pt, font=\scriptsize, text=blue!35!black,
           fill=white, inner sep=1pt] {$\kV_{mr}{=}\tfrac12$} (rA);
\end{tikzpicture}
\end{minipage}\hfill
\begin{minipage}{0.46\textwidth}
\centering
\textbf{(b) Horizontal ($\kH$)}\\[0.4em]
\begin{tikzpicture}[scale=1.25, transform shape]
  \node[draw, rounded corners, thick, minimum height=0.8cm, minimum width=1.5cm,
        text centered, font=\footnotesize,
        path picture={
          \fill[anticomp!30] (path picture bounding box.south west) rectangle ([xshift=0.75cm]path picture bounding box.south west |- path picture bounding box.north);
          \fill[policyblue!30] ([xshift=0.75cm]path picture bounding box.south west) rectangle (path picture bounding box.north east);
        }] (rA) {Refiner $r$};
  \node[draw, rounded corners, thick, minimum height=0.8cm, minimum width=1.5cm,
        text centered, font=\footnotesize, right=2.2cm of rA, fill=policyblue!30] (rB) {Refiner $j$};
  \node[draw, circle, thick, minimum size=0.9cm, text centered, font=\footnotesize,
        fill=anticomp!30, below=1.8cm of $(rA)!0.5!(rB)$] (m1) {Mine $m$};
  \node[font=\scriptsize, above=0.1cm of rA] {A: 50\% \; B: 50\%};
  \node[font=\scriptsize, above=0.1cm of rB] {B: 100\%};
  \node[font=\scriptsize, below=0.1cm of m1] {A: 100\%};
  \draw[-stealth, thick, red!70] (rA) to[bend left=20]
      node[midway, above=2pt, font=\scriptsize, text=red!70,
           fill=white, inner sep=1pt] {$\kH_{rj}{=}1$} (rB);
  \draw[-stealth, thick, red!40!black] (rB) to[bend left=20]
      node[midway, below=2pt, font=\scriptsize, text=red!40!black,
           fill=white, inner sep=1pt] {$\kH_{jr}{=}\tfrac12$} (rA);
\end{tikzpicture}
\end{minipage}

\vspace{0.6em}
{\footnotesize
\textcolor{anticomp!70}{$\blacksquare$}~Company~A \quad
\textcolor{policyblue!70}{$\blacksquare$}~Company~B \quad
\textcolor{blue!60!black}{\textbf{---}}~$\kV$ vertical \quad
\textcolor{red!70}{\textbf{---}}~$\kH$ horizontal}

\caption{The two ownership channels generated by one equity register.}
\label{fig:ownership_channels}
\begin{minipage}{0.95\textwidth}
\vspace{0.3em}
\footnotesize \textit{Notes:} Companies A and B own stakes in mine \mbox{$m$} and refiners \mbox{$r$} and \mbox{$j$}. The equity share \mbox{$\omega_{cf}$} records company \mbox{$c$}'s ownership of producer \mbox{$f$}. The profit-internalization weight \mbox{$\kappa_{fg}$} records how much producer \mbox{$f$} values producer \mbox{$g$}'s profit relative to its own; \mbox{$V$} denotes a vertical link and \mbox{$H$} a horizontal link. Both panels show the same ownership configuration:
$\omega_{Ar}=\omega_{Br}=\tfrac12$, $\omega_{Bj}=1$, and $\omega_{Am}=1$.
Under proportional control, equation~\mbox{\eqref{eq:kappa_def}} gives the profit-internalization weights
$\kV_{rm}=\kH_{rj}=1$ but
$\kV_{mr}=\kH_{jr}=\tfrac12$.
\end{minipage}
\end{figure}
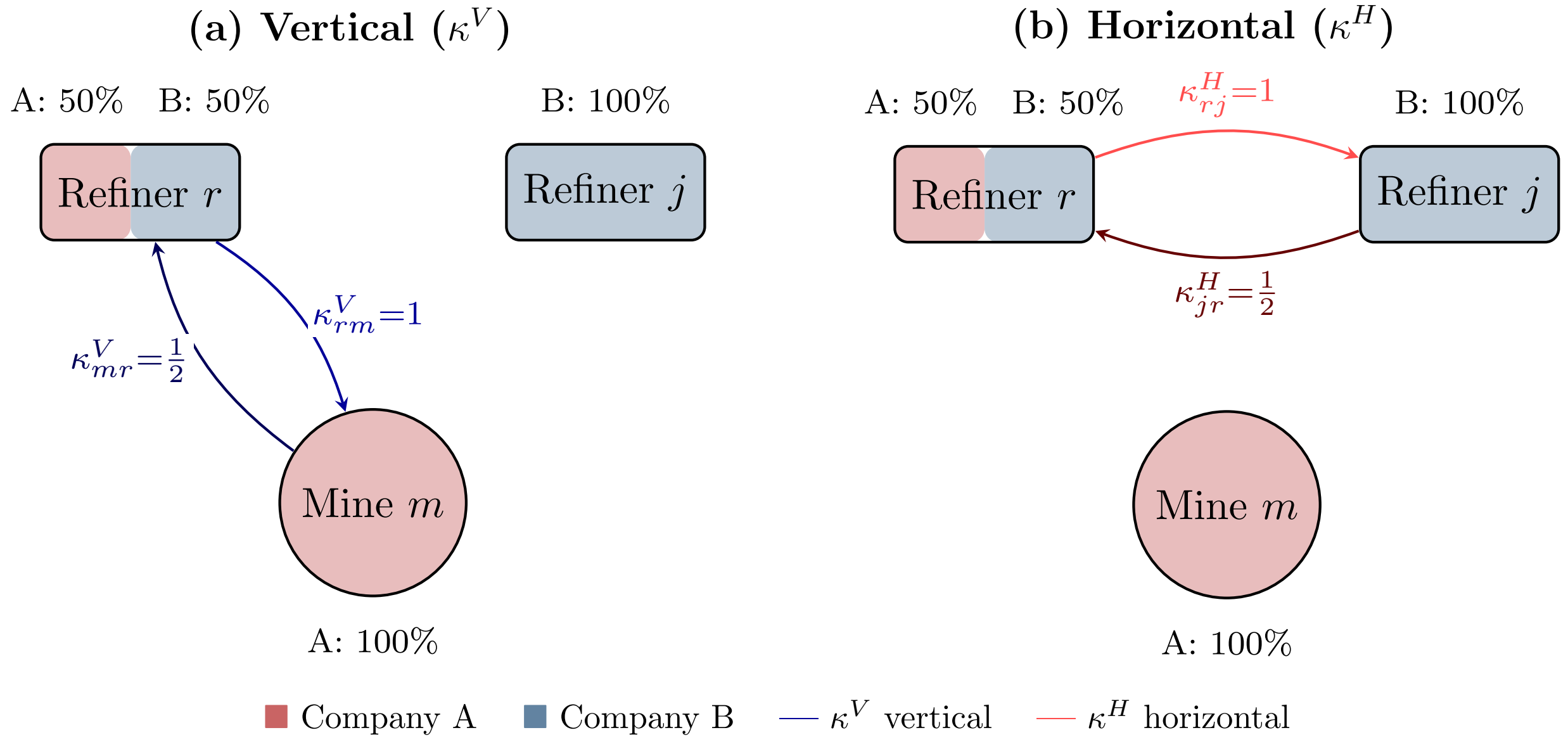

Figure~\ref{fig:ownership_channels} makes this joint-venture example concrete. Company A owns mine $m$ and half of refiner $r$; company B owns the other half of $r$ and all of rival refiner $j$. Refiner $r$ therefore values both the upstream profits of mine $m$ and the profits of its competitor $j$. The two panels show how the same ownership structure generates these vertical and horizontal profit interests.

\subsection[Market structure]{Market structure}
\label{sec:players_timing}

\subsubsection{Ore purchasing arrangements}
\label{sec:purchasing_arrangements}
Hard-rock refiners obtain ore through two channels. Purchases from affiliated mines using ownership-linked purchasing rights are called \textit{ownership-linked ore purchases}; purchases on the common ore market open to all hard-rock refiners are called \textit{spot ore purchases}. A refiner with ownership-linked rights can use both channels. Let $q_{rm}$ denote refiner $r$'s purchases from mine $m$ under ownership-linked rights.

Internalizing mining profits can make affiliated ore an attractive source of inputs. Purchasing rights limit the quantity refiners can obtain through this channel, so an ownership link does not provide unrestricted access to a mine's available production. I approximate these rights using equity-based ceilings shared across an owner's refiners. To track the use of a shared right, let $a^c_{rm}\ge0$ denote refiner $r$'s purchases from mine $m$ using owner $c$'s right, so $q_{rm}=\sum_c a^c_{rm}$.

\begin{samepage}
\begin{assumption}[Purchasing rights from equity]\label{ass:purchasing_rights}
\leavevmode
\begin{enumerate}[label=(\roman*),leftmargin=2.2em,itemsep=1pt]
\item \textit{Owner-level right.} Owner $c$'s purchases from mine $m$, summed across its eligible refiners, satisfy
\[
  \sum_r a^c_{rm}\le\omega_{cm}K_m.
\]
This right is shared across refiners. 
\item \textit{Allocation to a refiner.} I additionally restrict the amount owner $c$ can direct from mine $m$ to refiner $r$:
\[
  a^c_{rm}\le\min\{\omega_{cm},\omega_{cr}\}K_m.
\]
The refiner-equity restriction is additional to the owner-level right.
\end{enumerate}
\end{assumption}
\end{samepage}
Adding the limits in part~(ii) across owners gives a purchase ceiling for refiner $r$'s purchases from mine $m$:
\begin{equation}
  0\le q_{rm}\le\bar q_{rm}=\varphi_{rm}K_m,\qquad
  \varphi_{rm}=\sum_c\min\{\omega_{cm},\omega_{cr}\}.
  \label{eq:captive_entitlement}
\end{equation}
Part~(i) prevents an owner's mining stake from providing a separate entitlement at every affiliated refiner. Part~(ii) additionally limits how much of that entitlement can be directed to a partially owned refiner. Together, these ceilings bound the quantity of ore over which refiners can benefit from internalizing mining margins.

At Mount Holland, the joint venture agreement allocates production to partners in proportion to their ownership interests.\footnote{SQM, \textit{Mount Holland Lithium Project Technical Report Summary}, 25~April 2022, Section~16.7.} At Greenbushes, the Tianqi--IGO joint company elected to purchase less than its entitlement for the December 2023 quarter.\footnote{Each direct shareholder is entitled to purchase half of annual
production. See Tianqi Lithium, \textit{2022 Listing Prospectus}%
; IGO Limited, \textit{September 2023 Quarterly Activities Report},
30~October 2023. %
} These arrangements motivate ownership-linked purchase ceilings, with actual quantities chosen by refiners. These formulas approximate contractual rights; the refiner-level limit is an additional allocation assumption.

\begin{samepage}

\end{samepage}

\begin{samepage}
\begin{assumption}[Settlement of ownership-linked ore purchases]\label{ass:cup}
Ownership-linked ore purchases are made at the same price $w$ as spot ore purchases.
\end{assumption}
\end{samepage}
Disclosed arrangements link the prices of ownership-linked ore purchases to market prices. For example, Ganfeng reports purchases from Mount Marion at market prices. At Greenbushes, the offtake agreements specify that shareholders' purchase price for ore equals the prevailing price paid by third-party buyers when such sales occur. The model approximates these arrangements with a common price for ownership-linked and spot purchases, abstracting from contractual pricing lags, floors, and caps.\footnote{Ganfeng Lithium, \textit{2022 Annual Report} and \textit{2024 Annual Report} (A-share editions), related-party purchases; Tianqi Lithium, \textit{H1 2024 Interim Report}, Albemarle Off-take Agreement. Without third-party sales, the Greenbushes agreements use the last third-party price adjusted for changes in lithium carbonate prices. For examples of price floors and caps, see, respectively, Ganfeng's H-share prospectus, 27~September 2018, and Piedmont Lithium's press release, 2~August 2023. } Under this assumption, the vertical benefit comes from profit internalization: a refiner pays $w$ for ore but places weight $\kV_{rm}$ on the mining profit generated by its purchase.

\subsubsection[Producer decisions]{Producer decisions}

Each year is a separate play of a static, complete-information, two-stage game. Mines choose spot commitments before downstream producers choose sourcing and lithium output. This lets mines anticipate how additional ore changes downstream production and prices, and hence the profits of affiliated downstream producers.\footnote{Following the successive-oligopoly literature, I adopt upstream quantity leadership and downstream price taking in the ore market as maintained conduct assumptions \citep{S1988,GL1996}. } Equity, capacities, technologies, the demand state, and every producer's cost state are realized and common knowledge before either stage. Capacity, entry, and equity are predetermined within the year.

\begingroup
\clubpenalty=10000
\widowpenalty=10000
\paragraph[Upstream supply.]{Upstream supply.} Hard-rock mines simultaneously choose $s_m\in[0,K_m]$, the quantity of ore they commit to produce for the spot market. Each mine takes rival mines' commitments as given and anticipates how its own ore supply affects hard-rock refiners' sourcing decisions, downstream producers' lithium output, and the ore price.

\paragraph[Downstream sourcing and production.]{Downstream sourcing and production.} Given mine commitments, hard-rock refiners simultaneously choose lithium output and purchases under ownership-linked rights and on the spot market, taking the ore price $w$ and other strategic agents' output and sourcing choices as given. Brine producers choose lithium output in the same final-product market but do not purchase traded ore.
\endgroup

\phantomsection\label{ass:production_to_order}\label{ass:capacity}
Mines choose spot commitments $s_m$, while hard-rock refiners choose purchases $q_{rm}$ under ownership-linked rights. Mines fulfill these purchases subject to the capacity remaining after spot commitments, so total ore output is $x_m=s_m+\sum_r q_{rm}\le K_m.$

\subsubsection{Market clearing and equilibrium}

\begin{definition}[Equilibrium]\label{def:eq}
An equilibrium is a pure-strategy subgame-perfect Nash equilibrium of the two-stage game:
\begin{enumerate}
\item \textbf{Downstream choices.} Given mine commitments, strategic agents jointly choose their producers' output and sourcing to maximize their ownership-weighted objectives, taking other agents' choices and the ore price as given, subject to production capacities and sourcing constraints. Brine producers choose output in the same lithium market.
\item \textbf{Mine commitments.} Each mine's commitment is a best response to rival commitments, anticipating the downstream equilibrium following each possible commitment.
\item \textbf{Prices.} The lithium price $P(Q)$ is determined by total output through inverse demand. At a positive ore price $w$, spot ore purchases equal mines' spot commitments.
\end{enumerate}
I restrict attention to downstream equilibria supported by common shadow values on shared purchasing rights and mine capacity; \aref{app:sisters} states this selection formally.
\end{definition}

I characterize the game by working backward: downstream sourcing and output determine the response that upstream mines anticipate when choosing their spot commitments.

\subsection{Downstream characterization}
\label{sec:downstream_problem}

\subsubsection{Hard-rock refiners}
\label{sec:hardrock}
\label{sec:refiner}

\paragraph{The refiner's decision.}
A downstream strategic agent takes mines' spot commitments, the ore price $w$, and other strategic agents' outputs and purchases as given. A hard-rock refiner $r$ chooses output $q_r$, purchases $a^c_{rm}$ from mine $m$ under owner $c$'s right, and spot ore purchases $q_r^{\mathrm{spot}}$ to maximize $V_r$:
\begin{equation}
  \max_{q_r,\,\{a^c_{rm}\},\,q_r^{\mathrm{spot}}\ge0}\;
  \left\{\piR_r+\sum_{j\ne r}\kH_{rj}\piR_j
                     +\sum_m\kV_{rm}\piM_m\right\}.
  \label{eq:refiner_joint_problem}
\end{equation}
Total ore purchases must equal lithium output, $q_r$, in these units. The refiner can combine ownership-linked ore purchases and spot ore purchases, and output cannot exceed its capacity, $q_r\le K_r$. Ownership-linked ore purchases face two types of limits: the purchasing rights in Assumption~\ref{ass:purchasing_rights} and the supplying mine's capacity remaining after spot commitments. Other refiners' use of the same rights or mine capacity reduces the amount available to $r$. The full constraints are stated in \aref{app:ownerLP}.

Let $C_r^R(q_r)$ denote total refining cost, excluding ore purchases. Each unit of ore is priced at $w$, so the refiner's operating profit is
\[
  \piR_r=P(Q)q_r-wq_r-C_r^R(q_r).
\]

\paragraph{Ownership and ore sourcing.}
At fixed output, a change in sourcing leaves the refiner's ore payment and lithium revenue unchanged. It changes the profits of the supplying mines, however. A feasible additional tonne ordered from mine $m$ generates the mining margin $w-C_m^{M\prime}(x_m)$. The refiner internalizes that margin at weight $\kV_{rm}$, giving the effective marginal ore cost on route $(r,m)$
\begin{equation}
  c_{rm}=w-\kV_{rm}(w-C_m^{M\prime}(x_m))
        =(1-\kV_{rm})w+\kV_{rm}C_m^{M\prime}(x_m).
  \label{eq:crm}
\end{equation}
When $w>C_m^{M\prime}(x_m)$, ownership reduces this cost by the internalized share of the upstream profit. For $0<\kV_{rm}<1$, double marginalization is partially eliminated; at $\kV_{rm}=1$, the refiner values the ore at the mine's marginal cost. In the model, the refiner still pays $w$ for ore, but it internalizes part of the additional mining profit generated by its purchase. This financial interest in upstream profits is illustrated by Albemarle's accounts. The company reports that lower spodumene prices reduce its ore costs but also lower its earnings from Windfield, the holding company of the Greenbushes mine.\footnote{Albemarle Corporation, \textit{Form 10-K for 2025}, Item~7.}

\paragraph{Ownership and output.}
Expanding lithium output requires additional refining and ore. The marginal refining cost is $C_r^{R\prime}(q_r)$. Let $\lambda_r$ denote the ore-balance shadow value, which measures marginal sourcing cost. It accounts for profit internalization and sourcing constraints. The effective marginal production cost is therefore $C_r^{R\prime}(q_r)+\lambda_r$. At an interior output, the refiner's first-order condition is
\begin{equation}
  P(Q)-b(Q)\left(q_r+\sum_{j\ne r}\kH_{rj}q_j\right)
    =C_r^{R\prime}(q_r)+\lambda_r.
  \label{eq:foc_r}
\end{equation}
The left side is marginal revenue after accounting for the lithium-price reduction on own sales and on sales by other producers whose profits the refiner internalizes. The right side combines refining and effective ore costs. Vertical ownership affects the output incentive through $\lambda_r$; its mine-profit benefit is already included there. Horizontal ownership increases the revenue forgone through the lithium-price reduction and therefore discourages expansion, holding sourcing costs and other producers' outputs fixed.

To see how sourcing affects this condition at a given ore price, suppose additional ore comes from affiliated mine $m$ with purchasing limits and mine capacity slack. Then $\lambda_r=c_{rm}$. When $c_{rm}<w$, profit internalization lowers the ore cost of expanding output. Binding purchasing rights or mine capacity can add scarcity costs and limit this advantage. If the refiner already buys spot ore, $\lambda_r=w$: additional ore can be bought at the spot price. The refiner still internalizes mining profits on its ownership-linked ore purchases, but those purchases do not lower the cost of additional ore at that allocation. \aref{app:envelope} gives the full optimality conditions.

\subsubsection{Brine producers}
\label{sec:brine}

A brine producer combines extraction and refining at one site and purchases no traded ore. Let $B_r(q_r)$ denote its total production cost and $B_r'(q_r)$ its marginal production cost. It faces the same final-market tradeoff as a hard-rock refiner. At an interior output, its first-order condition is
\begin{equation}
  P(Q)-b(Q)\left(q_r+\sum_{j\ne r}\kH_{rj}q_j\right)=B_r'(q_r).
  \label{eq:foc_brine}
\end{equation}
Horizontal ownership operates through the same internalization of competitors' profits. At fixed output, the ore price does not enter brine marginal cost. Brine production nevertheless responds to changes in hard-rock output and the lithium price.

\paragraph{Downstream responses and upstream incentives.}
For given mine commitments, the downstream equilibrium specifies how much downstream producers produce, where hard-rock refiners obtain ore, and the price that clears the spot market. A change in spot supply can therefore affect downstream profits through ore costs and competition in the lithium market. Mines anticipate these effects when choosing spot supply and value them through their profit-internalization weights.

\subsection{Upstream characterization}
\label{sec:mine}

\paragraph{The mine's decision.}
For a commitment profile $\mathbf s=(s_m,\mathbf s_{-m})$, let $w(\mathbf s)$, $q_r(\mathbf s)$, and $q_{rk}(\mathbf s)$ denote the ore price, lithium output, and ownership-linked ore purchases from mine $k$ in the downstream equilibrium. Mine $m$ holds other mines' commitments $\mathbf s_{-m}$ fixed and chooses $s_m$ to maximize $V_m$:
\begin{equation}
  \max_{s_m\in[0,K_m]}\;
  \left\{
    \piM_m(\mathbf s)
    +\sum_{k\ne m}\kH_{mk}\piM_k(\mathbf s)
    +\sum_r\kV_{mr}\piR_r(\mathbf s)
  \right\}.
  \label{eq:mine_commitment_problem}
\end{equation}
The expression in braces is its continuation payoff $V_m(s_m;\mathbf s_{-m})$: every operating profit in~\eqref{eq:Vm} is evaluated at the downstream equilibrium induced by that commitment, including the lithium price $P(\sum_r q_r(\mathbf s))$. Ownership-linked ore purchases adjust within the capacity left by spot commitments, so each mine's total ore output is $x_k(\mathbf s)=s_k+\sum_r q_{rk}(\mathbf s)\le K_k$. Mine $m$'s operating payoff is therefore
\begin{equation}
  \piM_m(\mathbf s)
  =w(\mathbf s)x_m(\mathbf s)-C^M_m(x_m(\mathbf s))-\chi_m s_m,
  \label{eq:pi_mine}
\end{equation}
where $\chi_m$ is the \textit{spot-sale shock}, a mine-specific cost or benefit per unit of ore committed to the spot market. A positive value represents an additional cost of spot supply; a negative value represents a benefit. Possible sources include marketing or contract-servicing costs and the value of maintaining customer relationships. The shock is known to firms before choices are made but unobserved by the econometrician. The mine's optimality conditions are derived in \aref{app:mine_problem}.

\paragraph{Ownership and spot supply.}
In choosing spot supply, the mine accounts for its sales revenue, mining costs, and ownership-linked ore purchases. Ownership also makes it value effects on other producers' profits. The weight $\kV_{mr}$ determines how much it values changes in downstream producer $r$'s profits. If additional spot supply expands producer $r$'s competitors, their expansion puts downward pressure on the lithium price and on producer $r$'s earnings from existing output. For $\kV_{mr}>0$, this gives the mine an additional incentive to restrict spot supply: the input-foreclosure incentive.

For hard-rock refiners, additional spot supply can also lower ore costs. The reverse weight $\kV_{rm}$ affects refiner $r$'s effective ore costs and therefore its sourcing and output response, subject to purchasing rights and available capacity. Horizontal ownership among downstream producers also affects their output responses. The mine weighs the resulting downstream profit changes together with changes in its own mining profit.

Horizontal weights $\kH_{mk}$ make the mine also value changes in other mines' profits. When mine $m$ changes its spot supply, the ore price and ownership-linked purchases from other mines can change. Thus, even with their spot commitments fixed, other mines can experience changes in revenue and in the mining costs associated with their total output.

\section{Identification}\label{sec:identification}
\label{est-sec:overview}

I observe production, capacity, prices, equity ownership, and average variable costs (AVC) of extraction. Purchasing rights and profit-internalization weights are constructed from equity under the model's ownership mapping. I estimate demand parameters and technology-specific cost parameters for extraction and refining.

The extraction-cost equations relate observed AVC to utilization and other cost determinants, allowing me to construct marginal mining costs. Under maintained Cournot-with-ownership conduct, the first-order conditions for lithium production relate marginal revenue to extraction or sourcing costs and the remaining refining costs.

The equations share parameters. Demand elasticity enters both the demand equation and the first-order conditions for lithium production. Hard-rock mine costs enter refinery sourcing costs, while brine extraction parameters enter both the measured-cost equation and brine producers' first-order conditions. I estimate the equations jointly to impose these cross-equation restrictions. The following subsections explain the identifying variation and maintained restrictions for the demand and cost equations.

\subsection{Demand elasticity and inverse-demand slope}
\label{est-sec:demand}\label{id-sec:demand}

Regional refined-lithium demand is constant-elasticity (log--log) across regions $n$ and year $t$:
\begin{equation}
  \ln Q_{nt} \;=\; \zeta_n \;+\; \eta\,\ln P_t
  \;+\; \gamma_{\mathrm{GDP}}\ln\mathrm{GDP}_{nt} \;+\; \gamma_{\mathrm{Ni}}\ln\mathrm{Ni}_t
  \;+\; \gamma_{\mathrm{Oil}}\ln\mathrm{Oil}_t
  \;+\; \tau_D(t-\bar t) \;+\; \xi_{nt}.
\label{eq:ce_demand}
\end{equation}
Here $Q_{nt}$ is regional refined-lithium consumption, built as an accounting identity from final consumption of battery and non-battery products. $P_t$ is lithium price across regions. Regional intercepts absorb persistent demand differences. Real GDP controls for the scale of demand, nickel prices for substitution across cathode chemistries, and oil prices for the cost of the internal-combustion alternative and the energy cycle. The common linear trend, centered at the sample mean year $\bar t$, absorbs smooth secular demand growth.

Price responds to demand shocks, so I instrument it with $Z_t^D=\ln K^{\rm mine}_{t-1}$, the log of global hard-rock nameplate capacity in the previous year. More installed mining capacity expands the potential supply of ore, shifting refined-lithium supply and equilibrium prices. Exclusion is motivated by slow capacity adjustment: permitting and construction delays mean that capacity installed in the previous year reflects investment decisions made before current demand shocks. Investment can nevertheless respond to anticipated demand growth, which motivates including the demand controls and common linear trend. I assume that, conditional on these controls, lagged capacity is uncorrelated with unobserved demand shocks.

With a common price $P_t$ and elasticity, the regional demands in \eqref{eq:ce_demand} aggregate to
\begin{equation}
  Q_t(P_t)=A_tP_t^{\eta},\qquad \eta<0,
  \label{eq:model_ce_demand}
\end{equation}
where $ A_t= \exp\!\left[\gamma_{\mathrm{Ni}}\ln\mathrm{Ni}_t +\gamma_{\mathrm{Oil}}\ln\mathrm{Oil}_t +\tau_D(t-\bar t)\right] \sum_n\exp\!\left[\zeta_n +\gamma_{\mathrm{GDP}}\ln\mathrm{GDP}_{nt}+\xi_{nt}\right]. $ Regional intercepts, controls, the common trend, and demand shocks enter $A_t$ without changing the aggregate price elasticity. Holding these demand shifters fixed, inverse demand and its derivative are
\[
 P_t(Q)=\left(\frac{Q}{A_t}\right)^{1/\eta},\qquad
 \frac{\partial P_t(Q)}{\partial Q}
 =\frac{P_t(Q)}{\eta Q}.
\]
Evaluated at observed prices and quantities, the magnitude of the inverse-demand slope is $b_t(\eta)=P_t/(|\eta|Q_t)$.

\subsection{Hard-rock production costs}
\label{est-sec:engcost}

Hard-rock production links mining to refining through ore sourcing. I first explain how average variable costs map into marginal costs for both extraction technologies. I then show how hard-rock mining costs determine refiners' marginal ore-sourcing costs at observed production and prices, and how these sourcing costs enter the refiner's first-order condition for lithium production.

\subsubsection{From average to marginal mining cost}
\label{est-sec:avc-mc}
The dataset measures hard-rock mine extraction AVC and brine LiCl extraction AVC. Both cost blocks estimate how measured AVC changes with utilization and use the same mapping to marginal cost, with technology-specific parameters, controls, and utilization instruments. Let $u_{it}\equiv x_{it}/K_{it}$ denote utilization. The variable-cost primitive is
\[
  C_{it}(x)
  =
  \bigl(z_{it}'\theta+ \alpha u_{it} + \varepsilon_{it}\bigr)x_{it}.
\]
At the observed output $x_{it}$ this implies
\[
  \text{AVC}_{it}
  = z_{it}'\theta + \alpha\,u_{it} + \varepsilon_{it},
  \qquad
  \text{MC}_{it}
  = z_{it}'\theta + 2\alpha\,u_{it} + \varepsilon_{it},
\]
Within each extraction technology, I estimate a common utilization coefficient $\alpha$, while allowing heterogeneity in the marginal-cost slope with respect to output through differences in capacity: $\partial\mathrm{MC}_{it}/\partial x_{it}=2\alpha/K_{it}$. When $\alpha>0$, smaller-capacity sites have steeper marginal-cost curves because an additional tonne raises their utilization more.

The dataset reports average processing cost per unit, so each regression is fit directly to measured AVC,
\[
  \E\!\left[\varepsilon_{it}\cdot Z_{it}\right]=0,
  \qquad
  \varepsilon_{it} = \text{AVC}^{\rm data}_{it} - z_{it}'\theta - \alpha\,u_{it}.
\]
For hard-rock mines, utilization uses observed total ore output, including both spot sales and ownership-linked ore purchases. Estimating the mining-cost equation therefore requires neither solving for the split between these uses nor inverting the mine FOC. Utilization is endogenous because mines may produce more when their unobserved cost shocks are low. I therefore instrument for utilization. This use of observed average costs to recover marginal costs is related to \citet{DEH2026}.

\subsubsection{Mining costs}
\label{est-sec:avc}\label{id-sec:avc}
The sourcing and the mine's first-order conditions use marginal mining cost $C_{mt}^{M\prime}(x_{mt})$. The cost moment is the common measured-AVC moment above, $\E[\varepsilon_{mt}Z^M_{mt}]=0$, specialized to hard-rock mines. The controls $z_{mt}$ are a constant, log head grade, a non-spodumene indicator, log mining wage, and log electricity price. A mine's head grade measures the lithium content of its ore before processing, expressed as Li$_2$O content.

The instrument is the capacity of rival mines whose head grade exceeds that of the focal mine,
\[
  D^{\text{below}}_{mt} \;=\; \sum_{m'\neq m} K_{m't}\,\mathbf{1}\{g_{m't} > g_{mt}\}.
\]
The sum uses current nameplate capacity of producing mine cost units with observed grades strictly above that of $m$. Higher head grade tends to lower extraction cost per unit of lithium, conditional on other cost determinants. Capacity at higher-grade rivals can therefore displace demand for the focal mine's output and shift its utilization. Entry of new higher-grade capacity and expansion of incumbent rivals provide time variation, alongside cross-sectional differences in grade rankings. Exclusion is motivated by the slow adjustment of mining capacity and the geological basis of ore grade. Entry and capacity expansions require permitting, environmental review, and construction, limiting their response to current-year mining-cost shocks. Head grade reflects the deposit's geology.\footnote{In my annual head-grade data, grades are generally stable within mines, and relative grade rankings are generally stable among mines observed in consecutive years.} I therefore assume that, conditional on the focal mine's cost controls, higher-grade rival capacity is uncorrelated with its current mining-cost shock.

\subsubsection{Refining costs}
\label{est-sec:refinerfoc}
\label{est-sec:hrfoc}\label{id-sec:hrfoc}\label{id-sec:hr-corner}
\label{est-sec:lp}\label{id-sec:lambda}

I estimate hard-rock refining costs from the refiner's first-order condition. Let $\psi_{rt}(\theta_D)$ denote ownership-adjusted marginal revenue, evaluated at observed prices and quantities.
\label{app:hr_marginal_revenue}
\[
 \psi_{rt}(\eta)=P_t-b_t(\eta)
 \left[q_{rt}+\sum_{j\ne r}\kH_{rjt}(1-\tau^m_{jt})q_{jt}\right],
 \qquad b_t(\eta)=\frac{P_t}{|\eta|Q_t}.
\]
Here $\tau^m_{jt}$ denotes the marginal government take rate, set to zero for hard-rock refiners and defined for brine producers in Section~\ref{est-sec:brfoc}.

Following the first-order condition inversion in \citet{BH2014}, subtracting marginal sourcing cost from marginal revenue gives implied marginal refining cost:
\begin{equation}
 \underbrace{\psi_{rt}(\theta_D)-\lambda^{\mathrm{cond}}_{rt}(\theta_M)}_{\text{implied marginal refining cost}}
 =\alpha^{HR}_t+X_{rt}'\delta^{HR}+a^{HR}u_{rt}+\epsilon^{HR}_{rt}.
 \label{eq:hr-network-residual}
\end{equation}
Here $\lambda^{\mathrm{cond}}_{rt}(\theta_M)$ is the ore-balance shadow value conditional on observed production. I construct it by solving the sourcing program at observed production, ore prices, capacities, and ownership, using the candidate mine-cost parameters. This shadow value enters the refiner's first-order condition directly and updates with the mine-cost parameters during joint estimation.

Utilization is $u_{rt}=q_{rt}/K^{\rm eff}_{rt}$, and $\theta_D$ and $\theta_M$ denote demand and mine-cost parameters. The controls $X_{rt}$ are a mica indicator, log manufacturing wage, log electricity price, and an indicator for whether the refinery is grouped with sister plants. I also include year fixed effects $\alpha^{HR}_t$. The coefficient $a^{HR}$ measures how marginal refining cost changes with utilization. The residual $\epsilon^{HR}_{rt}$ captures variation not explained by the refining-cost specification.

Utilization is endogenous because refiners may produce more when their unobserved refining costs are low. I therefore instrument for utilization. The identifying restriction is
\begin{equation}
 \E\!\left[Z^{HR}_{rt}(\theta_{M})\,\epsilon^{HR}_{rt}
 \right]=0,
 \label{eq:hr-network-moment}
\end{equation}

For each rival refiner, I combine mine-cost shocks $\epsilon_{mt}$ using weights determined by its ownership links to mines. The instrument $z^{\rm net}_{rt}(\theta_M)$ averages these exposures across rival refiners, giving greater weight to refiners with larger nameplate capacity. It captures changes in rivals' sourcing costs that can shift their production and, consequently, demand for the focal refiner's output and its utilization. Exclusion is motivated by the distinct production processes and generally separate locations of hard-rock mining and refining. Mining shocks may reflect local extraction conditions, while refining shocks may reflect plant-specific conversion efficiency. Cost controls and year fixed effects account for observed input costs and common annual shocks. I assume that the weighted mining shocks affecting rivals are uncorrelated with the focal refiner's remaining refining-cost shock.

\subsection{Brine production costs}
\label{est-sec:brine}

Brine operations combine extraction and carbonation at the same site. I use a measured extraction-cost equation and a full-site first-order condition for lithium production. Their shared extraction-cost parameters link the two blocks, while the first-order condition also depends on demand elasticity. The blocks use common excluded instruments but impose restrictions on different residuals.

\subsubsection{Extraction costs}
\label{est-sec:brcost}\label{id-sec:brcost}
The dataset measures the cost of extracting and concentrating brine to produce lithium chloride (LiCl). I estimate this stage's average variable cost using the same specification as for hard-rock mining:
\[
 \mathrm{AVC}^{\rm LiCl}_{rt}
 =X^{BR\prime}_{rt}\delta^{BR}+a^{BR}u_{rt}+\varepsilon^{BR}_{rt}.
\]
The controls are a constant, log source-brine lithium concentration, a China indicator, log electricity price, and the prior-year Penman net-evaporation anomaly. Higher source-brine concentration means less brine must be pumped and concentrated per tonne of lithium. Electricity prices capture energy costs, while the China indicator captures persistent cost differences between Chinese and other operations. The evaporation anomaly measures whether weather conditions favor more or less water loss from evaporation ponds than usual at the site, accounting for both potential evaporation and rainfall. Higher values indicate conditions more favorable to concentrating brine in evaporation ponds; the one-year lag reflects the time needed for this process.

Utilization is endogenous because brine producers may produce more when their unobserved extraction costs are low. I instrument for utilization using the rival-capacity instruments described below.

\subsubsection[Full-site first-order condition]{Full-site first-order condition}
\label{est-sec:brfoc}\label{id-sec:brfoc}

With extraction and refining co-located, the brine producer's first-order condition equates net marginal revenue to full-site marginal cost. There is no traded ore input or separate sourcing decision.

I account for royalties, lease payments, and taxes linked to lithium prices by deducting them from revenue in the first-order condition. Let $G_{rt}(P)$ be the payment per tonne at price $P$. The producer retains $P-G_{rt}(P)$ per tonne, and a one-dollar price increase raises its net revenue per tonne by $1-G'_{rt}(P)$. Define the effective and marginal take rates at the observed price as
\[
 \tau^e_{rt}=\frac{G_{rt}(P_t)}{P_t},\qquad
 \tau^m_{rt}=G'_{rt}(P_t).
\]
The effective rate determines net revenue from an additional unit; the marginal rate determines the price effect on existing sales. Both rates are set to zero for hard-rock refiners. A brine affiliate's price effect is multiplied by $1-\tau^m_{jt}$ because its government payments also change with price. Under Cournot behavior with ownership links, a brine producer's net marginal revenue is
\[
 \psi^{BR}_{rt}(\eta)=(1-\tau^e_{rt})P_t
 -b_t(\eta)\left[(1-\tau^m_{rt})q_{rt}
 +\sum_{j\ne r}\kH_{rjt}(1-\tau^m_{jt})q_{jt}\right].
\]
The factor $(1-\tau^m_{jt})q_{jt}$ measures how much producer $j$'s net revenue changes with a one-dollar increase in the lithium price, holding its output fixed. The first-order condition separates extraction from carbonation costs:
\[
\psi^{BR}_{rt}
\;=\;
MC^{\rm LiCl,BR}_{rt}
+ MC^{\rm carb,BR}_{rt},
\qquad
MC^{\rm carb,BR}_{rt}
\;=\;
\text{soda}_{rt}+\phi_t+\tilde\omega^{BR}_{rt}.
\]
Where extraction AVC is observed, the constructed LiCl-stage marginal cost retains its realized cost component. The residual $\tilde\omega^{BR}_{rt}$ captures unobserved carbonation costs. The carbonation margin, the refining of LiCl into Li$_2$CO$_3$, adds no estimated cost column. Its dominant reagent, soda ash at $2.0$\,t/t, enters as a calibrated offset and a brine year fixed effect absorbs the residual between-year carbonation margin. The extraction-cost equation and the full-site first-order condition use the same utilization slope $a^{BR}$ and cost-control coefficients $\delta^{BR}$.

\subsubsection{Instruments and identifying restrictions}
I use the capacity of rival producers whose profits the brine producer does not internalize, and its interaction with log source-brine concentration, as excluded instruments:
\[
 z^{\rm riv}_{rt}=\sum_{k\ne r:\,\kH_{rkt}=0}K_{kt},\qquad
 Z^{BR}_{rt}=\begin{pmatrix}
 z^{\rm riv}_{rt}\\
 z^{\rm riv}_{rt}\ln c_r
 \end{pmatrix}.
\]
Here $K_{kt}$ is rival $k$'s nameplate capacity and $c_r$ is the site's source-brine concentration. Rival capacity changes competition in the refined-lithium market, shifting the demand facing the brine producer and its utilization. The interaction allows this response to differ with source concentration, which affects extraction costs.

The identifying restrictions are
\[
 \E[Z^{BR}_{rt}\varepsilon^{BR}_{rt}]=0,
 \qquad
 \E[Z^{BR}_{rt}\tilde\omega^{BR}_{rt}]=0.
\]
Both instruments must be uncorrelated with unobserved extraction costs in the AVC equation and with the remaining cost residual in the full-site first-order condition, after accounting for each equation's controls. Expanding rival capacity requires construction and commissioning. For brine operations, this involves developing wells, installing processing equipment, and expanding evaporation ponds. These investments take time, limiting capacity adjustment to unexpected current shocks. Source-brine concentration is largely determined by geology, and I account for its direct effect on extraction costs. I assume that the capacity of rivals whose profits the producer does not internalize, and its interaction with source concentration, are uncorrelated with the remaining cost shocks in each equation.

\FloatBarrier

\section[Estimation and model fit]{Estimation and model fit}
\label{sec:results}\label{sec:estimation}

This section describes the joint estimator and reports demand and cost estimates, then examines mine-output fit with the mine's first-order condition left out of estimation.

\subsection{Parameter estimates
 }
\label{est-sec:recspec}\label{est-sec:joint-main}

I estimate demand and costs jointly by generalized method of moments (GMM), combining the demand and mining-cost equations with the first-order conditions for lithium production in Section~\ref{sec:identification}.\footnote{I use two-step GMM with a diagonal weighting matrix, weighting each squared sample moment in the second step by the inverse of its year-clustered variance estimated from the first-step fit.} As mine-cost parameters change, I update refiners' sourcing costs and the network instrument, holding observed production, capacities, ownership, and ore prices fixed. I cluster standard errors by year, allowing shocks across producers and equations to be correlated within the same year. Table~\ref{est-tab:joint} reports the estimation results.

\begin{table}[H]
\centering
\caption{Joint-GMM estimates.}
\label{est-tab:joint}
\footnotesize
\begin{tabular}{@{}lrr@{}}
\toprule
Parameter & Estimate & SE \\
\midrule
\multicolumn{3}{@{}l}{\textit{Mine AVC}} \\
\DTRow{Utilization}{Mine AVC}{utilization}
\DTRow{Constant}{Mine AVC}{const}
\DTRow{Log head grade}{Mine AVC}{ln_grade}
\DTRow{Log wage}{Mine AVC}{ln_wage}
\DTRow{Log electricity price}{Mine AVC}{ln_energy}
\DTRow{Non-spodumene}{Mine AVC}{d_nonspod}
\addlinespace
\multicolumn{3}{@{}l}{\textit{Refiner MC}} \\
\DTRow{Utilization}{Refiner MC}{utilization}
\DTRow{Mica}{Refiner MC}{mica}
\DTRow{Log wage}{Refiner MC}{ln_wage}
\DTRow{Multiple-plant indicator}{Refiner MC}{multiplant}
\DTRow{Log electricity price}{Refiner MC}{ln_electricity}
\addlinespace
\multicolumn{3}{@{}l}{\textit{Brine AVC}} \\
\DTRow{Utilization}{Brine AVC}{utilization}
\DTRow{Constant}{Brine AVC}{const}
\DTRow{China}{Brine AVC}{China}
\DTRow{Log electricity price}{Brine AVC}{ln_electricity}
\DTRow{Log brine concentration}{Brine AVC}{ln_concentration}
\DTRow{Lagged evaporation anomaly (metres)}{Brine AVC}{lagged_evap_anomaly_metres}
\addlinespace
\multicolumn{3}{@{}l}{\textit{Demand}} \\
\DTRow[3]{Price elasticity}{Demand}{ln_price}
\DTRow[3]{Log GDP}{Demand}{ln_gdp}
\DTRow[3]{Log nickel price}{Demand}{ln_ni}
\DTRow[3]{Log oil price}{Demand}{ln_oil}
\DTRow[3]{Linear time trend}{Demand}{linear_time_trend}
\bottomrule
\end{tabular}
\begin{minipage}{0.95\textwidth}
\vspace{0.4em}
\footnotesize Notes: Costs are measured in  \$/t LCE.
The utilization coefficients measure changes in average variable cost for
mines and brine producers, and in marginal cost for hard-rock refiners.
Standard errors account for joint estimation and are clustered by year.
\end{minipage}
\end{table}

The joint demand elasticity is $\widehat\eta=\DTEst[3]{Demand}{ln_price}$ (SE \DTSE[3]{Demand}{ln_price}). It is informed by both regional demand and lithium producers' first-order conditions. Cost curvature governs how marginal costs change as production expands. The estimated hard-rock refining-cost slope is \mbox{\DTEst{Refiner MC}{utilization}}. A ten-percentage-point increase in utilization raises marginal refining cost by approximately \$\DTScaled{0.1}{Refiner MC}{utilization}{} per tonne LCE. The mine and brine AVC slopes imply marginal-cost slopes of \$\DTScaled{2}{Mine AVC}{utilization}{} and \$\DTScaled{2}{Brine AVC}{utilization}{} per tonne LCE for a unit increase in utilization. The table also reports the geology, technology, and factor-price coefficients.

\subsection{Mine-output fit}
\label{est-sec:minefit}

I assess whether the estimated model can reproduce observed mine production. The mine optimality condition is excluded from estimation, so estimation does not require observed mine output to satisfy that condition.

\begin{figure}[!htbp]
\centering
\includegraphics[width=0.6\textwidth]{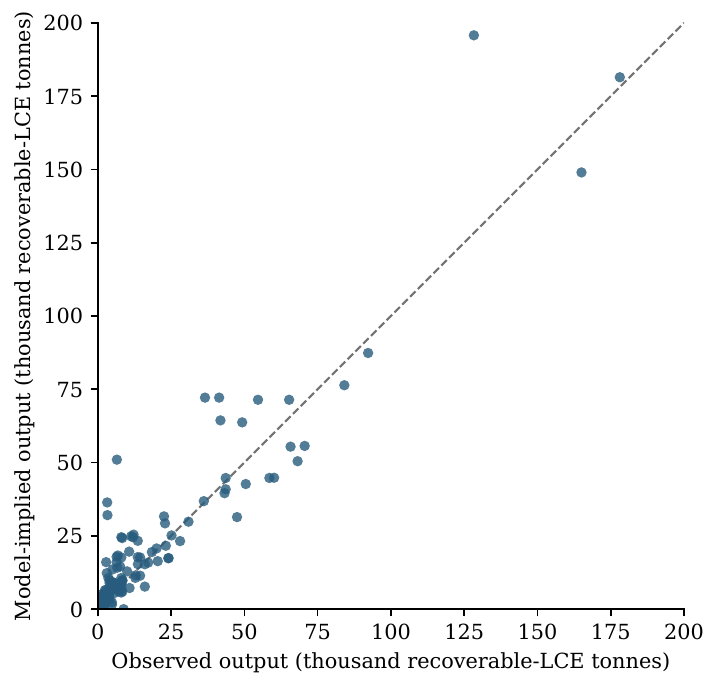}
\caption{Observed and model-implied mine output.}
\label{est-fig:minefit-retained}
\begin{minipage}{0.95\textwidth}
\footnotesize\smallskip
\textit{Notes:} Each point represents an active hard-rock mine in a given year.
The sample contains $N=\MFERNAll{}$ observations over \MFERYears{}.
Observed and model-implied quantities are measured in thousands of
recoverable-LCE tonnes; the dashed $45^\circ$ line indicates exact agreement.
Prediction $R^2$ is \MFERMainRSquared{}, RMSE is \MFERMainRMSE{}, and aggregate
mine output is overpredicted by \MFERMainAggregate\%.
\end{minipage}
\end{figure}

I hold estimated demand parameters, ownership, and capacities fixed. I retain the estimated cost coefficients and recovered mine extraction-cost shocks. On the refinery side, I retain the estimated cost coefficients and recovered cost shocks. These cost schedules remain fixed as production and sourcing adjust. I set mine spot-sale shocks $\chi_{mt}$ to zero, so no additional incentive to sell ore on the spot market is fitted to reproduce observed mine output.

Each mine chooses its spot commitment taking other mines' commitments as given and accounting for its own profit and the other producers' profits weighted by ownership, as described in~\eqref{eq:mine_commitment_problem}. For each candidate commitment profile, I recompute downstream production and sourcing, the market-clearing ore price, and the lithium price. Ownership-linked purchases adjust within the mine capacity remaining after spot commitments. Predicted mine output is therefore $\widehat x_{mt}=\widehat s_{mt}+\sum_r\widehat q_{rmt},$ the sum of the mine's spot commitment and ownership-linked purchases from that mine.

Figure~\ref{est-fig:minefit-retained} compares predicted and observed mine output for \MFERNAll{} mine-years over \MFERYears{}. Prediction $R^2$ is \MFERMainRSquared{}, and RMSE is \MFERMainRMSE{} thousand recoverable-LCE tonnes. Aggregate mine output is overpredicted by \MFERMainAggregate\%. This comparison assesses the production choices implied by the mine optimality condition, conditional on the estimated parameters and recovered cost shocks.
\FloatBarrier

\section{Counterfactuals}
\label{sec:counterfactuals}

Ownership connects producers' incentives across production stages and among competitors. A refiner's stake in a mine can lower its effective input cost because it shares in the mining margin on affiliated ore purchases. The mine may also restrict spot ore supply to protect affiliated refiners' profits when additional supply would expand their rivals' production. Horizontal ownership links among final lithium producers can discourage expansion by making each producer internalize the decline in competing producers' profits. These relationships overlap in the lithium industry, where firms hold interests in hard-rock mines, refineries, and integrated brine operations. I use counterfactual changes in ownership and trade to examine how these incentives jointly shape supply, sourcing, production, and prices.

Motivated by governments' concerns about reliable access to mineral inputs, I compare two interventions: divestiture of Chinese stakes in overseas lithium producers and a ban on ore exports to China. Divestiture transfers Chinese stakes in overseas hard-rock mines and integrated brine operations to separate owners in the host countries with no existing interests in lithium production, while leaving trade open. The export ban retains observed ownership but prohibits overseas ore deliveries to Chinese refineries through both spot and affiliated routes. Divestiture changes the incentives to supply and source ore while allowing Chinese refiners to continue purchasing it through international trade; the ban directly restricts their access to overseas ore. The comparison shows how these policies differ in their effects on ore and lithium supply.

The exercises are conducted for \CFCurYear{} using the demand and cost estimates reported in Table~\ref{est-tab:joint}. Taking each intervention as given, I solve for firms' responses under ownership-weighted profit maximization. Within each comparison, capacity, entry, estimated parameters, and royalties remain fixed, while mines' spot commitments, refinery sourcing and production, and prices adjust. Each policy effect is measured against a re-solved observed-ownership benchmark, with mine spot-sale shocks set to zero in both economies. These comparisons are informative about short- to medium-run effects, given the time required to develop and commission new mining and refining capacity.

In Section~\ref{sub:cf-q1}, I examine the supply and sourcing responses to divestiture under observed ownership portfolios. Section~\mbox{\ref{sub:cf-q2}} assesses how these responses depend on downstream competition by repeating the divestiture under pooled Chinese refining interests. Trade remains open in both exercises. Finally, Section~\mbox{\ref{sub:cf-q3}} compares divestiture with the ore export ban.

\subsection{How does cross-border ownership shape supply and sourcing?}
\label{sub:cf-q1}

Cross-border ownership can shape mines' incentives to supply ore to the spot market, refiners' sourcing decisions, and competition in the final lithium market. To study these effects, I compare observed ownership with the transfer of Chinese stakes in overseas mines and the integrated brine operation to independent local owners. Trade remains open. Table~\mbox{\ref{tab:cf-q1-outcomes}} reports sourcing, production, and prices under observed ownership and divestiture.

\begin{table}[H]
\centering
\caption{Effects of divestiture under observed portfolios, \CFCurYear.}
\label{tab:cf-q1-outcomes}
\small
\begin{tabular}{@{}lrrrr@{}}
\toprule
Outcome & \shortstack{Observed\\ownership} & Divested & Change & Change (\%) \\
\midrule
Spot ore purchases (t LCE) & $\CFZeroOneSpotF$ & $\CFZeroOneSpotD$ & $+\CFZeroOneSpotChangeAbs$ & $+\CFZeroOneSpotPctAbs$ \\
Ownership-linked ore purchases (t LCE) & $\CFZeroOneCaptiveF$ & $\CFZeroOneCaptiveD$ & $-\CFZeroOneCaptiveChangeAbs$ & $-\CFZeroOneCaptivePctAbs$ \\
Ore price (USD/t LCE) & $\CFZeroOneOreF$ & $\CFZeroOneOreD$ & $\CFZeroOneOreChange$ & $\CFZeroOneOrePctChange$ \\
\midrule
Lithium output (t LCE) & $\CFZeroOneOutputF$ & $\CFZeroOneOutputD$ & $+\CFZeroOneOutputChangeAbs$ & $+\CFZeroOneOutputPctAbs$ \\
\quad Hard-rock & $\CFZeroOneHardrockF$ & $\CFZeroOneHardrockD$ & $-\CFZeroOneHardrockChangeAbs$ & $-\CFZeroOneHardrockPctAbs$ \\
\quad Brine & $\CFZeroOneBrineF$ & $\CFZeroOneBrineD$ & $+\CFZeroOneBrineChangeAbs$ & $+\CFZeroOneBrinePctAbs$ \\
Lithium price (USD/t LCE) & $\CFZeroOnePriceCNF$ & $\CFZeroOnePriceCND$ & $-\CFZeroOnePriceCNChangeAbs$ & $\CFZeroOnePriceCNPctChange$ \\
\bottomrule
\end{tabular}

\vspace{0.4em}
\begin{minipage}{0.95\textwidth}
\footnotesize \textit{Notes:} Each ownership column reports the solution obtained by re-solving the model under that ownership structure. 
The change column reports divested ownership minus observed ownership; percentage changes are relative to the observed-ownership level. Totals and changes are calculated before rounding. 
\end{minipage}
\end{table}

Divestiture substantially changes ore sourcing while leaving aggregate lithium supply nearly unchanged. Spot ore purchases rise by \CFZeroOneSpotPctAbs\%, while ownership-linked purchases fall by \CFZeroOneCaptivePctAbs\%. Yet the ore price rises and total ore use declines, indicating that the shift toward spot trading does not bring cheaper inputs or greater overall ore use. Total lithium output increases by only \CFZeroOneOutputPctAbs\%, as brine expansion largely offsets the contraction in hard-rock refining, and the lithium price falls by \CFZeroOnePricePctAbs\%.

Understanding this response requires accounting for the incentives attached to ownership. When mining margins are positive, affiliated refiners benefit from sharing in their suppliers' mining profits. Mines, in turn, consider how spot supply affects affiliated refiners' profits. Divestiture removes these ownership links, changing both refiners' sourcing incentives and mines' incentives to supply spot ore. The increase in spot purchases and the higher ore price reflect the joint adjustment of supply, sourcing, and production.

The small change in aggregate lithium supply also depends on the horizontal interests removed by divestiture. Brine operations compete with hard-rock refineries but require no traded ore. Divesting the Chinese stake in the affected brine operation reduces its interest in competing refiners' profits, strengthening its incentive to expand production. In the reported solutions, aggregate brine output rises despite the lower lithium price and largely offsets the contraction in hard-rock refining. Nearly unchanged final lithium supply thus masks substantial changes in sourcing and production across the two technologies.

\paragraph{Policy implications.}
Divestiture shifts ore purchases toward the spot market, while brine expansion largely offsets the decline in hard-rock refining and leaves global lithium supply nearly unchanged. For governments concerned about domestic input access, the key distinction is between transferring ownership of a mine and securing its output for domestic users. Chinese refiners remain able to purchase spot ore, and the equilibrium ore price rises. The ownership transfer therefore changes the incentives governing supply and sourcing without ensuring that domestic refiners obtain more ore or pay less for it.

\subsection{How does downstream competition shape the effects of cross-border ownership?}
\label{sub:cf-q2}

Downstream competition can shape how changes in ownership affect production and demand for ore. To study this interaction, I compare the divestiture in Section~\mbox{\ref{sub:cf-q1}} under observed portfolios and pooled Chinese refining interests. Trade remains open in both environments.

Pooling combines Chinese owners' refining stakes, including their holdings in foreign brine operations, into a single portfolio used to construct the weights on competing refiners' profits. Within each ownership scenario, it leaves weights on mining profits, mines' weights on refining profits, and affiliated purchasing rights unchanged. This represents a hypothetical change in incentives, not an estimate of observed coordination. I measure the effect of divestiture separately in each environment, comparing each divested economy with its corresponding pre-divestiture benchmark. Table~\mbox{\ref{tab:cf-coordination}} reports sourcing, production, and prices before and after divestiture under pooled Chinese refining interests.

\begin{table}[H]
\centering
\caption{Effects of divestiture under pooled Chinese refining interests, \CFCurYear{}.}
\label{tab:cf-coordination}
\small
\begin{tabular}{@{}lrrrr@{}}
\toprule
Outcome & \shortstack{Before\\divestiture} & \shortstack{After\\divestiture} & Change & Change (\%) \\
\midrule
Spot ore purchases (t LCE) & $\CFUpdatedTwoPoolSpotF$ & $\CFUpdatedTwoPoolSpotD$ & $\CFUpdatedTwoPoolSpotChange$ & $\CFUpdatedTwoPoolSpotPct$ \\
Ownership-linked ore purchases (t LCE) & $\CFUpdatedTwoPoolCaptiveF$ & $\CFUpdatedTwoPoolCaptiveD$ & $\CFUpdatedTwoPoolCaptiveChange$ & $\CFUpdatedTwoPoolCaptivePct$ \\
Ore price (USD/t LCE) & $\CFUpdatedTwoPoolOreF$ & $\CFUpdatedTwoPoolOreD$ & $\CFUpdatedTwoPoolOreChange$ & $\CFUpdatedTwoPoolOrePct$ \\
\midrule
Lithium output (t LCE) & $\CFUpdatedTwoPoolOutputF$ & $\CFUpdatedTwoPoolOutputD$ & $\CFUpdatedTwoPoolOutputChange$ & $\CFUpdatedTwoPoolOutputPct$ \\
\quad Hard-rock & $\CFUpdatedTwoPoolHardrockF$ & $\CFUpdatedTwoPoolHardrockD$ & $\CFUpdatedTwoPoolHardrockChange$ & $\CFUpdatedTwoPoolHardrockPct$ \\
\quad Brine & $\CFUpdatedTwoPoolBrineF$ & $\CFUpdatedTwoPoolBrineD$ & $\CFUpdatedTwoPoolBrineChange$ & $\CFUpdatedTwoPoolBrinePct$ \\
Lithium price (USD/t LCE) & $\CFUpdatedTwoPoolPriceF$ & $\CFUpdatedTwoPoolPriceD$ & $\CFUpdatedTwoPoolPriceChange$ & $\CFUpdatedTwoPoolPricePct$ \\
\bottomrule
\end{tabular}

\vspace{0.4em}
\begin{minipage}{0.95\textwidth}
\footnotesize \textit{Notes:} Each column reports the solution before or after divestiture under pooled Chinese refining interests. Pooling includes Chinese owners' foreign brine holdings; downstream profit weights are recomputed using the ownership interests remaining after divestiture. Changes are calculated as after minus before divestiture; percentage changes are relative to the pre-divestiture level under pooling. Totals and changes are calculated before rounding.
\end{minipage}
\end{table}

Divestiture produces a larger increase in lithium supply under pooled refining interests, but spot ore purchases decline. In this comparison, lithium output rises by \CFUpdatedTwoPoolOutputPctAbs\%, compared with \CFUpdatedTwoBaseOutputPctAbs\% under observed portfolios (Table~\mbox{\ref{tab:cf-q1-outcomes}}). Spot purchases fall by \CFUpdatedTwoPoolSpotPctAbs\%, whereas they rise by \CFUpdatedTwoBaseSpotPctAbs\% under observed portfolios. Each change is measured against the pre-divestiture economy in the same competitive environment.

Understanding these differences requires accounting for how downstream ownership interests affect production incentives. Pooling broadens the competing refining interests whose profits the affected brine operation internalizes. Holding other producers' output fixed, expanding brine production lowers the lithium price and reduces competing refiners' profits. Under pooling, the brine operation places greater weight on these profit losses, strengthening its incentive to restrain production. Divestiture removes the Chinese ownership link carrying these interests. In the reported solutions, aggregate brine output consequently expands more under pooling, more than offsetting the larger contraction in hard-rock refining.

The change in production composition also explains why greater lithium supply can accompany lower spot ore purchases. Brine expansion requires no traded ore, while the contraction in hard-rock refining reduces total ore use. Under observed portfolios, the decline in ownership-linked purchases exceeds the decline in total ore use, so spot purchases increase. Under pooled interests, the reduction in total ore use exceeds the reduction in ownership-linked purchases, so spot purchases also decline. The sourcing response therefore depends on how divestiture changes the composition of lithium production and the resulting demand for ore.

\paragraph{Policy implications.}
For governments concerned about domestic input access, evaluating divestiture requires understanding competition among final lithium producers. The same ownership transfer increases spot ore purchases under observed portfolios but reduces them under pooled refining interests. Removing ownership links therefore need not increase spot trading. Under pooling, the larger increase in lithium supply comes from brine expansion, even as hard-rock refining and ore use contract. The consequences for input access thus depend on downstream production incentives as well as the ownership links removed by the policy.

\FloatBarrier

\subsection{How do ownership and trade restrictions affect the lithium supply chain?}
\label{sub:cf-q3}

Restrictions on ore trade change which refiners can obtain inputs and how mines and refiners adjust production. Divestiture changes ownership incentives but leaves Chinese refiners able to purchase ore through international trade. I therefore compare the divestiture in Section~\mbox{\ref{sub:cf-q1}} with a ban on ore exports to China to examine how these interventions affect ore access, production, and prices along the lithium supply chain.

Divestiture transfers the targeted stakes while leaving trade open. The ban retains ownership but prohibits ore deliveries from the rest of the world to China through both spot and affiliated routes. Reverse ore trade, from China to the rest of the world, remains permitted. Ore trade among countries outside China remains permitted, so the ban does not reserve each country's ore for its own refiners. Both policies are measured against the same re-solved economy with observed ownership and no ban.

All three economies use the same demand and cost estimates and zero mine spot-sale shocks. However, unlike Sections~\ref{sub:cf-q1} and~\ref{sub:cf-q2}, this comparison imposes a common positive floor on mine marginal costs to prevent negative marginal costs at low output under the ore export ban.\footnote{I set the floor to the lowest marginal mining cost across hard-rock mine-year observations. } The benchmark is re-solved with the same floor, so the comparisons within this question hold the cost calibration fixed. Table~\mbox{\ref{tab:cf-trade}} reports the ore-export-ban outcomes relative to the benchmark with observed ownership and no ban.\footnote{Re-solving the observed-ownership and divestiture economies with the common cost floor reproduces the results in Table~\ref{tab:cf-q1-outcomes} to the reported precision. That table therefore also provides the divestiture comparison for this subsection.}

\begin{table}[H]
\centering
\caption{Effects of a ban on ore exports to China, \CFCurYear.}
\label{tab:cf-trade}
\small
\begin{tabular}{@{}lrrrr@{}}
\toprule
Outcome & No ban & \shortstack{Ore export\\ban} & Change & Change (\%) \\
\midrule
Spot ore purchases (t LCE) & $\CFThreeBanTableSpotF$ & $\CFThreeBanTableSpotB$ & $\CFThreeBanTableSpotChange$ & $\CFThreeBanTableSpotPct$ \\
Ownership-linked ore purchases (t LCE) & $\CFThreeBanTableCaptiveF$ & $\CFThreeBanTableCaptiveB$ & $\CFThreeBanTableCaptiveChange$ & $\CFThreeBanTableCaptivePct$ \\
Ore price (USD/t LCE) & & & & \\
\quad China & $\CFThreeBanTableOreChinaF$ & $\CFThreeBanTableOreChinaB$ & $\CFThreeBanTableOreChinaChange$ & $\CFThreeBanTableOreChinaPct$ \\
\quad RoW & $\CFThreeBanTableOreRoWF$ & $\CFThreeBanTableOreRoWB$ & $\CFThreeBanTableOreRoWChange$ & $\CFThreeBanTableOreRoWPct$ \\
\midrule
Lithium output (t LCE) & $\CFThreeBanTableOutputF$ & $\CFThreeBanTableOutputB$ & $\CFThreeBanTableOutputChange$ & $\CFThreeBanTableOutputPct$ \\
\quad Hard-rock & $\CFThreeBanTableHardrockF$ & $\CFThreeBanTableHardrockB$ & $\CFThreeBanTableHardrockChange$ & $\CFThreeBanTableHardrockPct$ \\
\quad Brine & $\CFThreeBanTableBrineF$ & $\CFThreeBanTableBrineB$ & $\CFThreeBanTableBrineChange$ & $\CFThreeBanTableBrinePct$ \\
Lithium price (USD/t LCE) & $\CFThreeBanTablePriceF$ & $\CFThreeBanTablePriceB$ & $\CFThreeBanTablePriceChange$ & $\CFThreeBanTablePricePct$ \\
\bottomrule
\end{tabular}

\vspace{0.4em}
\begin{minipage}{0.95\textwidth}
\footnotesize \textit{Notes:} Each scenario column reports the solution obtained by re-solving the model with observed ownership, either without or with the ore export ban. Both scenarios impose the same positive floor on mine marginal costs. The divestiture comparison matches Table~\mbox{\ref{tab:cf-q1-outcomes}} to the reported precision and is not repeated here. The change column reports ore export ban minus no ban; percentage changes are relative to the no-ban level. Totals and changes are calculated before rounding. Ore purchases and lithium output are global totals. Ore prices differ between China and the rest of the world (RoW), while refined lithium has a common world price.
\end{minipage}
\end{table}

Divestiture leaves global lithium supply nearly unchanged, raising output by \CFUpdatedThreeDivestOutputPctAbs\%. The ore export ban instead reduces output by \CFUpdatedThreeBanOutputPctAbs\% and raises the world lithium price by \CFUpdatedThreeBanPricePctAbs\%. Ore prices move in opposite directions across regions: they rise in China and fall outside China. These changes accompany a sharp contraction in hard-rock refining, partially offset by an expansion in brine production.

The ban prevents Chinese refiners from obtaining overseas ore through either spot or affiliated purchases. Their sourcing options narrow, and the Chinese ore price rises. Outside China, refiners face cheaper ore and a higher lithium price, strengthening their incentive to expand production. However, hard-rock refineries outside China reach capacity in the reported solution, limiting their ability to process ore that can no longer reach Chinese plants. Production outside China replaces only \CFUpdatedThreeBanReplacementPct\% of China's lost output.

Brine production provides another source of adjustment because it requires no traded ore. The higher lithium price encourages brine expansion, which partially offsets the decline in hard-rock production. This response is insufficient to restore global lithium supply. Cheaper ore outside China therefore coexists with a higher price for final lithium: the refineries permitted to purchase that ore cannot expand enough to replace the lost production.

\pagebreak[3]
\paragraph{Policy implications.}
{For governments concerned about domestic input access, the comparison distinguishes changing ownership from restricting foreign access to ore. Divestiture shifts purchases toward the spot market while allowing Chinese refiners to continue purchasing ore through international trade, with little change in global lithium supply. The export ban prevents overseas ore from reaching Chinese refineries and lowers ore prices outside China, but existing refining capacity elsewhere cannot fully replace the lost production. The ban therefore makes ore cheaper outside China while reducing global lithium supply and raising the price of final lithium. The supply cost of the restriction depends on whether the refineries permitted to receive the ore have sufficient capacity to process it.}

\FloatBarrier

\section{Conclusion}
\label{sec:conclusion}

Using global data on mines and refineries, I estimate a model of ownership and production to study how cross-border ownership shapes ore supply, refinery sourcing, lithium production, and prices. I compare divestiture of Chinese stakes in overseas mines and integrated brine operations with a restriction on ore exports to China, taking these geopolitically motivated interventions as given and holding capacity and entry fixed. Ownership restrictions change affiliated sourcing advantages, mines' incentives to supply spot ore, and producers' interests in competing firms' profits. Their supply effects therefore depend on the vertical and horizontal interests connecting producers.

Under observed portfolios, divestiture increases spot ore purchases and the ore price while leaving aggregate lithium output nearly unchanged, as brine expansion largely offsets the contraction in hard-rock refining. The comparison under hypothetical pooling of Chinese refining interests suggests a larger output response to divestiture, driven primarily by brine expansion, even as spot ore purchases fall. The same divestiture can therefore increase or decrease spot purchases depending on downstream competitive incentives. The ore-export restriction substantially reduces global lithium supply and raises lithium prices, despite lowering ore prices outside China.

The counterfactuals reveal a constraint on policies intended to reduce dependence on Chinese refining. Divestiture changes who owns mineral resources, but continued trade allows Chinese refiners to purchase ore through the spot market. Removing Chinese ownership therefore does not itself ensure that the ore is redirected toward domestic production. An export restriction directly prevents overseas ore shipments to China, but existing refining capacity elsewhere cannot fully replace the lost production. At existing capacities, changing ownership leaves the route to Chinese refining open, while closing that route imposes a substantial cost in lithium supply.

For governments seeking reliable access to critical minerals, these findings show why ownership incentives and refining capacity must be considered together. Ownership and trade restrictions can alter access to inputs, but they cannot substitute for the refining capacity needed to use those inputs. Investment in refining capacity outside China could therefore change the usefulness of these policies, allowing more redirected ore to support lithium production. In future work, I plan to examine how investment in refining capacity outside China changes the effectiveness of ownership and trade restrictions in securing mineral supply chains.

\clearpage
\bibliographystyle{econ}
\bibliography{references}

\clearpage
\appendix
\section*{Appendix}
\section{Data descriptives}
\label{app:descriptive_evidence}
\label{app:descriptive_figures}

\begin{table}[H]
\centering
\setstretch{1.05}
\caption{Summary statistics, lithium supply-chain panel, 2015--2024.}
\label{tab:sumstats}
\footnotesize
\setlength{\tabcolsep}{6pt}
\begin{tabular}{l rrrrr}
\toprule
 & Mean & Std.\ dev.\ & $p_{25}$ & Median & $p_{75}$ \\
\midrule
\multicolumn{6}{l}{\textit{Panel A: Hard-Rock Mines}} \\
\addlinespace[3pt]
\quad Production (kt LCE) & 20.5 & 34.7 & 2.9 & 8.0 & 19.5 \\
\quad Nameplate capacity (kt LCE) & 34.5 & 44.0 & 9.1 & 20.0 & 40.0 \\
\quad Utilization rate (\%) & 56.0 & 30.4 & 28.3 & 54.4 & 81.9 \\
\quad Variable cost (\$/t LCE) & 3,409 & 1,668 & 2,125 & 2,985 & 4,550 \\
\quad Li$_2$O head grade (\%) & 1.07 & 0.54 & 0.55 & 1.06 & 1.40 \\
\quad Mining wage (USD/month) & 3,066 & 3,225 & 789 & 1,028 & 7,636 \\
\quad Industrial electricity (US cents/kWh) & 11.55 & 4.21 & 9.18 & 9.69 & 11.93 \\
\midrule
\multicolumn{6}{l}{\textit{Panel B: Chemical Producers ($r \in \mathcal{R}$)}} \\
\addlinespace[2pt]
\multicolumn{6}{l}{\quad \textsc{Hard Rock Refiners} ($r \in \mathcal{R}^{HR}$)} \\
\addlinespace[3pt]
\quad Production (kt LCE) & 9.6 & 8.8 & 3.6 & 7.0 & 14.0 \\
\quad Nameplate capacity (kt LCE) & 20.8 & 18.6 & 9.4 & 15.0 & 25.0 \\
\quad Utilization rate (\%) & 55.8 & 29.9 & 30.0 & 55.4 & 80.0 \\
\quad Manufacturing wage (USD/month) & 982 & 399 & 801 & 930 & 1,112 \\
\quad Industrial electricity (US cents/kWh) & 10.13 & 1.93 & 9.18 & 9.65 & 10.61 \\
\addlinespace[12pt]
\multicolumn{6}{l}{\quad \textsc{Integrated Brine Producers} ($r \in \mathcal{R}^{BR}$)} \\
\addlinespace[3pt]
\quad Chemical production (kt LCE) & 16.4 & 24.5 & 3.0 & 6.8 & 21.7 \\
\quad Nameplate capacity (kt LCE) & 36.9 & 54.0 & 10.0 & 19.0 & 40.0 \\
\quad Utilization rate (\%) & 47.4 & 24.4 & 27.7 & 50.0 & 65.5 \\
\quad LiCl-stage variable cost (\$/t LCE) & 5,436 & 1,557 & 4,121 & 5,550 & 6,134 \\
\quad Source-brine Li concentration (mg/L Li$^{+}$) & 745.9 & 621.9 & 310.0 & 632.0 & 728.0 \\
\quad Industrial electricity (US cents/kWh) & 7.92 & 2.73 & 5.94 & 7.42 & 9.63 \\
\quad Prior-year net-evaporation anomaly (mm) & -5.5 & 106.5 & -36.4 & 3.2 & 61.9 \\
\midrule
\multicolumn{6}{l}{\textit{Panel C: Prices and Demand (Annual)}} \\
\quad Li$_2$CO$_3$ CIF Asia (\$/t) & 19,305 & 18,298 & 9,419 & 11,835 & 15,046 \\
\quad Spodumene FOB Australia (\$/t) & 1,331 & 1,398 & 600 & 729 & 898 \\
\quad Global lithium demand (kt LCE) & 495.3 & 331.7 & 245.6 & 338.2 & 698.9 \\
\bottomrule
\end{tabular}\par
\begin{minipage}{0.95\textwidth}
\vspace{0.4em}
\footnotesize \textit{Notes:} Production and capacity are expressed in lithium
carbonate equivalent (LCE). Utilization is annual production divided by
nameplate capacity.
\end{minipage}
\end{table}
\FloatBarrier

\clearpage
\section{Model}
\label{app:derivations}

This section derives the optimality conditions from the ownership-weighted objectives. It first states the decision units and feasible set, then derives refiner optimality and the downstream equilibrium responses that an upstream mine anticipates when choosing its spot commitment. The mine derivative accounts for changes in sourcing allocation and in which constraints bind as a mine changes its spot commitment and the downstream equilibrium adjusts.

\subsection{Objectives, decision units, and feasibility}
\label{app:refiner_implementation}

\subsubsection{Decision units and objectives}
\label{app:company_agg}

Each plant is a separate decision-maker unless it has the same owners, holding the same respective shares, as another plant. Plants with identical ownership make decisions jointly and form a single strategic agent. In the estimation sample, these groups consist of wholly owned sister plants. Each agent chooses output for its plants and ore purchases for its hard-rock plants. Each plant retains its own technology, production costs, and capacity. Ore-purchase rights depend on owners' equity stakes in mines and hard-rock refineries.

Let $\mathcal P_a$ be strategic agent $a$'s plants and $Q_a=\sum_{r\in\mathcal P_a}q_r$. Their common ownership vector implies a unit profit-internalization weight between any two plants in the portfolio and identical weights on every outside producer. Writing these common weights as $\kH_{aj}$ and $\kV_{am}$, the objective is
\begin{equation}
  V_a=\sum_{r\in\mathcal P_a}\piR_r
      +\sum_{j\notin\mathcal P_a}\kH_{aj}\piR_j
      +\sum_m\kV_{am}\piM_m.
  \label{eq:company_objective}
\end{equation}
A singleton portfolio recovers~\eqref{eq:Vr}.

As in the main text, $q_{rm}=\sum_c a^c_{rm}$ denotes ownership-linked ore purchases and $q_r^{\mathrm{spot}}$ denotes spot ore purchases. Define
\[
  A_m=\sum_{r\in\mathcal R^{\mathrm{HR}}}q_{rm},\qquad
  x_m=s_m+A_m,\qquad Q=\sum_rq_r.
\]
Let $C_m^M(x_m)$ be a mine's total mining cost, $C_r^R(q_r)$ a hard-rock plant's total refining cost, excluding ore purchases, and $B_r(q_r)$ a brine plant's total production cost. The mine's marginal cost is $C_m^{M\prime}(x_m)$. Ore flows $q_{rm}$, $a^c_{rm}$, and $q_r^{\mathrm{spot}}$ are zero for brine plants, so sums of these flows implicitly run over hard-rock plants. The operating payoffs are
\begin{align}
  \piM_m&=wx_m-C_m^M(x_m)-\chi_ms_m,
       &&(m\in\mathcal M),\label{eq:B_pim}\\
  \piR_r&=Pq_r-wq_r-C_r^R(q_r)
       &&(r\in\mathcal R^{\mathrm{HR}}),\label{eq:B_pir}\\
  \piR_r&=Pq_r-B_r(q_r)
       &&(r\in\mathcal R^{\mathrm{BR}}).\label{eq:B_pibr}
\end{align}

\subsubsection{Owner-indexed purchasing rights}
\label{app:ownerLP}

For given spot commitments, ore purchases must be nonnegative and satisfy the following constraints, where \mbox{$r$} indexes hard-rock refineries throughout:
\begin{align}
  \sum_{m,c}a^c_{rm}+q_r^{\mathrm{spot}}&=q_r
      &&\forall r,\label{eq:A_demand}\\
  a^c_{rm}&\le\min(\omega_{cm},\omega_{cr})K_m
      &&\forall r,c,m,\label{eq:A_chan}\\
  \sum_r a^c_{rm}&\le\omega_{cm}K_m
      &&\forall c,m,\label{eq:A_claim}\\
  \sum_{r,c}a^c_{rm}&\le K_m-s_m
      &&\forall m.\label{eq:A_out}
\end{align}
All plants also satisfy $0\le q_r\le K_r$. The channel cap limits what an owner can direct through a particular plant. The owner-right row prevents the same mine stake from being used independently at several plants. The last row reserves capacity for the mine's spot commitment. Primitive equity shares determine these rights; the directional profit-internalization weights $\kV$ determine the value assigned to mine profits.

When a strategic agent deviates unilaterally, other agents' outputs and purchases are fixed. Their purchases are subtracted from the right-hand sides of the shared rows~\eqref{eq:A_claim}--\eqref{eq:A_out}. A variational equilibrium uses common multipliers on these shared rows.

\subsection{Downstream optimality}
\label{app:refiner_problem}

\subsubsection{Conditional sourcing cost}

Fix all plants' outputs, mines' spot commitments $\mathbf s$, the ore price $w$, and other strategic agents' purchases. Each hard-rock plant then requires a fixed quantity of ore. Because every ore tonne is purchased at $w$, a feasible change in sourcing leaves all downstream producers' operating profits unchanged. It changes mine operating profits, which the purchasing agent values through its vertical profit-internalization weights.

For strategic agent $a$, let
\[
  F_{am}=\sum_{r\in\mathcal P_a\cap\mathcal R^{\mathrm{HR}}}q_{rm},
  \qquad x_{m,-a}=s_m+\sum_{j\notin\mathcal P_a}q_{jm}.
\]
Here $F_{am}$ aggregates ownership-linked purchases from mine $m$ across all of agent $a$'s hard-rock plants, and $x_{m,-a}$ is mine output excluding those purchases. Thus $x_m=x_{m,-a}+F_{am}$, with $x_{m,-a}$ fixed in this conditional problem. Relative to $F_{am}=0$, these purchases generate mine revenue $wF_{am}$ and additional mining cost $C_m^M(x_{m,-a}+F_{am})-C_m^M(x_{m,-a})$. The agent's ore payment less the mine profit it internalizes is therefore
\[
  \underbrace{wF_{am}}_{\text{ore payment}}
  -\underbrace{\kV_{am}\left[wF_{am}
    -\{C_m^M(x_{m,-a}+F_{am})-C_m^M(x_{m,-a})\}\right]}_{\text{internalized mine profit}}.
\]
Adding spot purchases and summing over mines gives the effective sourcing cost below. With outputs, the ore price, mine commitments, and rivals' purchases fixed, maximizing the agent's objective over its sourcing choices is equivalent to minimizing this cost. Payoff terms that do not change with the agent's sourcing choices are omitted because they do not affect its optimal purchases.
\begin{equation}
\begin{split}
  \Psi_a(\mathbf a_a,\mathbf q_a^{\mathrm{spot}})
  ={}&w\sum_{r\in\mathcal P_a\cap\mathcal R^{\mathrm{HR}}}q_r^{\mathrm{spot}}\\
     &+\sum_m\left[(1-\kV_{am})wF_{am}
       +\kV_{am}\{C_m^M(x_{m,-a}+F_{am})-C_m^M(x_{m,-a})\}\right].
\end{split}
\label{eq:exact_procurement}
\end{equation}
Consequently, for a route used by plant $r\in\mathcal P_a$,
\begin{equation}
  \frac{\partial\Psi_a}{\partial a^c_{rm}}
  =(1-\kV_{am})w+\kV_{am}C_m^{M\prime}(x_m)=c_{rm},
  \qquad \frac{\partial\Psi_a}{\partial q_r^{\mathrm{spot}}}=w.
  \label{eq:procurement_gradient}
\end{equation}
This is the effective marginal ore cost on route $(r,m)$ in~\eqref{eq:crm}.

For an ownership-linked marginal route with slack sourcing constraints and \mbox{$0<\kV_{am}<1$}, the saving relative to spot ore is \mbox{$\kV_{am}(w-C_m^{M\prime}(x_m))$}. Holding mine marginal cost fixed, this saving rises with \mbox{$w$} at rate \mbox{$\kV_{am}$}, while the effective marginal sourcing cost rises at rate \mbox{$1-\kV_{am}$}. When \mbox{$w>C_m^{M\prime}(x_m)$}, a higher ore price therefore widens the refiner's cost advantage over spot buyers while also raising its own effective sourcing cost.

\begin{samepage}
\begin{lemma}[Sourcing separation]\label{lem:coupling}
Given outputs, mines' spot commitments, the spot ore price \mbox{$w$}, and other strategic agents' purchases, a feasible change in a strategic agent's sourcing leaves all downstream producers' operating profits unchanged and changes only mine operating profits. Horizontal profit-internalization weights therefore do not enter this conditional sourcing choice; vertical weights enter through the effective marginal ore costs $c_{rm}$.
\end{lemma}
\end{samepage}

\begin{samepage}
Hold the ore price, mine commitments, and other agents' outputs and ore purchases fixed. For given plant outputs $\mathbf q_a$, let $C_a^*(\mathbf q_a)$ be the minimum effective sourcing cost obtained by choosing agent $a$'s ownership-linked and spot purchases:
\begin{equation}
  C_a^*(\mathbf q_a)
  =\min_{\mathbf a_a,\mathbf q_a^{\mathrm{spot}}\ge0}
     \{\Psi_a:\ \text{constraints \eqref{eq:A_demand}--\eqref{eq:A_out}}\}.
  \label{eq:A_Cstar}
\end{equation}
\end{samepage}%
Here $\Psi_a$ is defined in~\eqref{eq:exact_procurement}, and the shared purchasing constraints account for other agents' fixed purchases. When $\mathbf q_a$ changes, the agent reoptimizes its purchases to meet the resulting ore requirements, keeping the ore price, mine commitments, and other agents' choices fixed.

\subsubsection{Sourcing KKT conditions and the marginal source}

With convex mining costs and nonnegative profit-internalization weights, the sourcing problem is convex. Let $\lambda_r$ be the unrestricted ore-balance multiplier, and let $\tilde\nu^c_{rm}$, $\nu_{cm}$, and $\mu_m$ be the nonnegative multipliers on the channel, owner-right, and remaining-capacity rows. I write the conditions using the common owner-right and mine-capacity multipliers imposed by the variational equilibrium selection in \aref{app:downstream_response}. An individual agent's optimization problem alone does not require these multipliers to be equal across agents. The strategic agent's sourcing Lagrangian is
\begin{equation}
\begin{split}
  \mathcal L_a={}&\Psi_a
  -\sum_{r\in\mathcal P_a\cap\mathcal R^{\mathrm{HR}}}
       \lambda_r\left(\sum_{m,c}a^c_{rm}+q_r^{\mathrm{spot}}-q_r\right)\\
  &+\sum_{r\in\mathcal P_a\cap\mathcal R^{\mathrm{HR}},c,m}
       \tilde\nu^c_{rm}\{a^c_{rm}-\min(\omega_{cm},\omega_{cr})K_m\}\\
  &+\sum_{c,m}\nu_{cm}\left(\sum_r a^c_{rm}-\omega_{cm}K_m\right)
   +\sum_m\mu_m\left(\sum_{r,c}a^c_{rm}-K_m+s_m\right).
\end{split}
\label{eq:A_lagrangian}
\end{equation}
Rival purchases in the last two sums are constants in strategic agent $a$'s problem. Stationarity and complementary slackness give
\begin{equation}
  q_r^{\mathrm{spot}}\ge0,\qquad w-\lambda_r\ge0,\qquad
  q_r^{\mathrm{spot}}(w-\lambda_r)=0,
  \label{eq:A_stat_spot}
\end{equation}
\begin{equation}
  a^c_{rm}\ge0,\quad
  c_{rm}+\tilde\nu^c_{rm}+\nu_{cm}+\mu_m-\lambda_r\ge0,\quad
  a^c_{rm}\bigl(c_{rm}+\tilde\nu^c_{rm}+\nu_{cm}+\mu_m-\lambda_r\bigr)=0,
\label{eq:A_stat_route}
\end{equation}
together with feasibility and complementary slackness on all capacity rows. Thus spot ore purchases imply $\lambda_r=w$. A positive ownership-linked route satisfies $\lambda_r=c_{rm}+\tilde\nu^c_{rm}+\nu_{cm}+\mu_m$. For $q_r>0$, at least one source is positive, so
\begin{equation}
  \lambda_r
   =\min\!\left(w,\min_{(m,c)\text{ eligible for }r}
       \{c_{rm}+\tilde\nu^c_{rm}+\nu_{cm}+\mu_m\}\right).
  \label{eq:A_lambda}
\end{equation}
Equivalently, the combined owner-right and channel rent on route $(r,m)$ is
\[
  \nu_{rm}=\min_{c\text{ eligible}}(\nu_{cm}+\tilde\nu^c_{rm}).
\]

\begin{lemma}[Marginal source and shadow value]\label{lem:merit}
For a plant with positive output, the sourcing KKT conditions imply
\begin{equation}
  \lambda_r=
  \begin{cases}
    c_{r\hat m},&\text{if an ownership-linked route is used with zero scarcity rents},\\
    w,&\text{if spot ore purchases are positive}.
  \end{cases}
  \label{eq:lambda_merit}
\end{equation}
Here $\hat m$ denotes a mine supplying plant $r$ through a positive ownership-linked purchase with zero channel, owner-right, and mine-capacity multipliers. Its effective marginal cost equals $\lambda_r$; routes with lower effective marginal ore costs may be constrained. When a sourcing constraint binds, its multiplier enters the ownership-linked route's effective marginal cost as in~\eqref{eq:A_stat_route}.
\end{lemma}

\subsubsection[Output optimality conditions]{Output optimality conditions}
\label{app:envelope}

The output optimality conditions use the ore-balance multipliers \mbox{$\lambda_r$} from the sourcing KKT system.\footnote{When the minimized sourcing cost \mbox{$C_a^*$} is differentiable, \mbox{$\lambda_r$} equals its derivative with respect to \mbox{$q_r$}, holding other outputs fixed. At kinks, the same vector of shadow values must satisfy both the sourcing and output optimality conditions.} The minimized sourcing cost $C_a^*(\mathbf q_a)$ is the agent's ore expenditure net of the additional mine profits it internalizes from its purchases. After optimizing sourcing, agent $a$ therefore chooses its plant outputs to maximize
\begin{equation}
\begin{split}
  P(Q)Q_a-C_a^*(\mathbf q_a)
  &-\sum_{r\in\mathcal P_a\cap\mathcal R^{\mathrm{HR}}}C_r^R(q_r)\\
  &-\sum_{r\in\mathcal P_a\cap\mathcal R^{\mathrm{BR}}}B_r(q_r)
  +\sum_{j\notin\mathcal P_a}\kH_{aj}\piR_j.
\end{split}
\label{eq:Vr_nested}
\end{equation}
Using \mbox{$P'(Q)=-b(Q)$}, the output optimality conditions for hard-rock refineries are:
\begin{align}
  \zeta_r&\equiv P(Q)-b(Q)\left(Q_a+\sum_{j\notin\mathcal P_a}\kH_{aj}q_j\right)
                    -C_r^{R\prime}(q_r)-\lambda_r
             =\nu_r^q-\eta_r^q,\label{eq:refiner_output_kkt}\\
  \nu_r^q(K_r-q_r)&=0,\qquad \eta_r^q q_r=0,
             \qquad \nu_r^q,\eta_r^q\ge0.\notag
\end{align}
For a brine plant, replace $C_r^{R\prime}+\lambda_r$ by $B_r'$. Agent $a$ chooses output and sourcing jointly for all plants in $\mathcal P_a$, and the condition above applies to each plant in this joint problem. At an interior output choice, ownership-adjusted marginal revenue equals marginal refining cost plus marginal sourcing cost.

\paragraph{Allocation of output across plants.}
An agent operating several plants chooses both total output and its allocation across plants. When choosing total output, it accounts for the price reduction on sales from all its plants. For a given total output $Q_a$, strategic agent $a$ allocates production to minimize total plant costs plus $C_a^*(\mathbf q_a)$, subject to $\sum_{r\in\mathcal P_a}q_r=Q_a$ and plant capacities. The positive fitted utilization slope makes hard-rock refining costs strictly convex. With strictly convex plant costs, the cost-minimizing allocation of a given total output across plants is unique, although tied sourcing flows need not be.

\subsection{Downstream equilibrium and the ore-price response}
\label{app:downstream_response}

\subsubsection{Shared constraints and equilibrium selection}
\label{app:sisters}

For each commitment vector $\mathbf s$, the downstream equilibrium determines strategic agents' output and sourcing choices, mines' marginal extraction costs evaluated at the resulting mine output, and an ore price $w$ that clears the spot market. Each strategic agent takes $w$ and the other agents' choices as given. Shared purchasing constraints make each agent's feasible choices depend on other agents' purchases. I select a variational equilibrium by requiring common multipliers $\nu_{cm}$ and $\mu_m$ on these constraints, while each agent has its own balance, channel, and output multipliers.

Each strategic agent minimizes its own sourcing cost, taking other agents' purchases as given. Ownership differences mean that agents value changes in mining profits differently, even when they face common shadow values on shared constraints. I therefore solve their individual sourcing optimality conditions jointly.

\subsubsection{Downstream equilibrium conditions}
\label{app:equilibrium_conditions}

For a given mine commitment vector $\mathbf s$, the downstream outcome consists of $(P,w,\mathbf q,\mathbf a,\mathbf q^{\mathrm{spot}})$ and supporting multipliers. Given mines' spot commitments, a downstream equilibrium must satisfy the following conditions, as specified in Definition~\mbox{\ref{def:eq}}:
\begin{enumerate}
\item \textbf{Optimal choices and common shadow values.} Given the ore price $w$, mine commitments $\mathbf s$, and other agents' choices, each strategic agent jointly chooses output and sourcing to maximize \eqref{eq:company_objective} subject to the constraints above. Its output and sourcing conditions must hold simultaneously. Under the variational selection, agents use the same multipliers $\nu_{cm}$ and $\mu_m$ on the shared purchasing-right and mine-capacity constraints, respectively; each agent retains its own balance, channel, and output multipliers.
\item \textbf{Cost consistency.} The sourcing conditions use marginal mining cost $C_m^{M\prime}(x_m)$ evaluated at the induced mine output $x_m=s_m+\sum_{r,c}a^c_{rm}$.
  \item \textbf{Market clearing.} The prices satisfy
  \begin{equation}
    \begin{gathered}
    P=P(Q),\qquad w\ge0,\qquad
    \sum_m s_m-\sum_r q_r^{\mathrm{spot}}\ge0,\\
    w\left(\sum_m s_m-\sum_r q_r^{\mathrm{spot}}\right)=0.
    \end{gathered}
    \label{eq:ore_clearing}
  \end{equation}
\end{enumerate}

\subsubsection{Ore-price response to mine commitments}
\label{app:stage2_derivative}
\label{mp-app:lambda}

I examine how downstream output, sourcing, and prices respond to a change in mine $m$'s spot commitment, holding rival commitments fixed. An increase in mine $m$'s spot commitment adds to spot supply, but can also change refiners' spot demand by affecting mining costs and the availability of ownership-linked ore. To distinguish these effects from the price adjustment, let $D(w,\mathbf s)$ denote total spot demand when downstream agents choose output and sourcing at a given ore price and commitment vector.

The derivative $D_w$ measures the response of spot demand to the ore price, holding commitments fixed. The derivative $D_{s_m}$ measures its response to mine $m$'s commitment, holding the ore price and rival commitments fixed. At a positive equilibrium ore price, spot demand equals spot supply: $D(w,\mathbf s)=\sum_k s_k$. Along a differentiable equilibrium path,
\[
  D_w\frac{dw}{ds_m}+D_{s_m}=1.
\]
Thus, provided $D_w\ne0$,
\begin{equation}
  \frac{dw}{ds_m}=\frac{1-D_{s_m}}{D_w}=-\Lambda_m,
  \qquad \Lambda_m\equiv-\frac{dw}{ds_m}.
  \label{eq:B_dw}
\end{equation}
If the additional commitment leaves spot demand unchanged at the original price ($D_{s_m}=0$), the ore price must adjust to induce purchases of the extra supply. With downward-sloping spot demand ($D_w<0$), the price falls, and the magnitude of the price response is $\Lambda_m=1/|D_w|$.

However, spot demand need not remain fixed. Even with spare mine capacity, an additional commitment can change mining marginal costs and refiners' sourcing choices. When mine capacity is fully used, an additional spot tonne can instead displace one tonne of ownership-linked purchases. If refiners replace that tonne with a spot purchase and nothing else changes, then $D_{s_m}=1$: spot supply and demand increase equally, so the ore price does not change.

These derivative formulas apply where the downstream equilibrium varies smoothly with the mine's commitment. At a kink, the responses to an increase and a decrease may differ, so I use the relevant one-sided derivative when it exists. Alternatively, I evaluate a change in commitment directly: holding rival commitments fixed, I solve for the new downstream equilibrium and compare the mine's ownership-weighted payoff with its original payoff. This direct comparison also applies when derivatives are unavailable.

\subsection{Mine optimality and interpretation}
\label{app:mine_problem}

\subsubsection{The objective and the spot-sale shock}
\label{app:merchant_adjustment}
\label{app:mine_roadmap}

Mine $m$ chooses its spot commitment $s_m$, holding rival commitments $\mathbf s_{-m}$ fixed and anticipating downstream responses. Let \mbox{$z^*(s_m,\mathbf s_{-m})$} denote downstream equilibrium output, sourcing, and prices for each feasible commitment profile. Full equilibrium in Definition~\ref{def:eq} requires each mine to solve
\begin{equation}
  \max_{0\le s_m\le K_m}\quad
  \piM_m(z^*,\mathbf s)
  +\sum_{k\ne m}\kH_{mk}\piM_k(z^*,\mathbf s)
  +\sum_r\kV_{mr}\piR_r(z^*,\mathbf s).
  \label{eq:mine_continuation_objective_app}
\end{equation}
The three terms are the mine's own profit, the profits of other mines it internalizes, and the downstream profits it internalizes. Each is evaluated at the downstream outcome induced by $s_m$. Rival mine output $x_k=s_k+A_k$ can therefore change even though $s_k$ is fixed, because ownership-linked orders $A_k$ respond. The directional weight $\kV_{mr}$ measures how much mine $m$ values refiner $r$'s profit; the reverse weight $\kV_{rm}$ measures how much the refiner values the mine's profit.

The spot-sale shock $\chi_m$, defined in Section~\ref{sec:mine}, enters the mine's payoff as $-\chi_ms_m$. A positive value adds a cost per unit committed to spot sales; a negative value adds a benefit. Given mine commitments, this term is constant with respect to refiners' sourcing and output choices. It therefore affects downstream decisions through the mine's chosen commitment and the resulting capacity and price responses, with no direct term in the refiner's optimality conditions.

\subsubsection[Mine payoff derivative and ownership incentives]{Mine payoff derivative and ownership incentives}
\label{app:mine_deriv}
\label{mp-app:equiv}

I derive the change in mine \mbox{$m$}'s ownership-weighted payoff from a marginal increase in its spot commitment \mbox{$s_m$}, holding rival commitments fixed. The derivation follows the downstream equilibrium as it varies smoothly with \mbox{$s_m$}, with binding constraints unchanged.

\paragraph{Operating-profit responses.}
All quantity, price, and payoff derivatives below are total derivatives along the downstream equilibrium with respect to \mbox{$s_m$}. Primes on cost functions denote derivatives with respect to their own arguments. In particular, $\frac{dw}{ds_m}=-\Lambda_m$, $\frac{dP}{ds_m}=-b(Q)\frac{dQ}{ds_m}$, and $\frac{dx_k}{ds_m}=\mathbf 1\{k=m\}+\frac{dA_k}{ds_m}$. Direct differentiation of the operating payoffs gives
\begin{align}
  \frac{d\piM_m}{ds_m}&=(w-C_m^{M\prime}(x_m))\left(1+\frac{dA_m}{ds_m}\right)-\Lambda_m x_m-\chi_m,
      \label{eq:B_dpim}\\
  \frac{d\piM_k}{ds_m}&=(w-C_k^{M\prime}(x_k))\frac{dA_k}{ds_m}-\Lambda_m x_k\quad(k\ne m),
      \label{eq:B_dpik}\\
  \frac{d\piR_r}{ds_m}&=-b(Q)\frac{dQ}{ds_m}q_r+\Lambda_m q_r
      \notag\\
      &\quad +(P-w-C_r^{R\prime})\frac{dq_r}{ds_m}
      \quad(r\in\mathcal R^{\mathrm{HR}}),\label{eq:B_dpir_raw}\\
  \frac{d\piR_r}{ds_m}&=-b(Q)\frac{dQ}{ds_m}q_r+(P-B_r')\frac{dq_r}{ds_m}
      \quad(r\in\mathcal R^{\mathrm{BR}}).\label{eq:B_dpibr_raw}
\end{align}
For mine $m$, total production changes by $1+dA_m/ds_m$: the additional spot commitment can also change ownership-linked orders. The mine earns its mining margin on this production change, experiences the ore-price effect on all its sales, and incurs the marginal spot-sale shock $\chi_m$. Other mines' production changes only through their ownership-linked orders.

\paragraph{Using downstream optimality.}
I use the refiner's output condition to rewrite the terms involving its own output response in~\eqref{eq:B_dpir_raw}. Let $\zeta_r=\nu_r^q-\eta_r^q$ denote the contribution of the output-bound multipliers in~\eqref{eq:refiner_output_kkt}, where $\nu_r^q$ is the multiplier on capacity ($q_r\le K_r$) and $\eta_r^q$ is the multiplier on nonnegativity ($q_r\ge0$). Both are zero at an interior output, so $\zeta_r=0$ there. Rearranging that condition for a hard-rock plant gives
\begin{equation}
  P-w-C_r^{R\prime}-b(Q)q_r
    =b(Q)\sum_{j\ne r}\kH_{rj}q_j+(\lambda_r-w)+\zeta_r.
  \label{eq:B_refiner_residual}
\end{equation}
\phantomsection\label{rem:residual}%
First split the total output response into the plant's own response and those of other plants: $dQ/ds_m=dq_r/ds_m+\sum_{j\ne r}dq_j/ds_m$. Combining the own-output terms in~\eqref{eq:B_dpir_raw} gives the coefficient $P-w-C_r^{R\prime}-b(Q)q_r$ on $dq_r/ds_m$. Substituting~\eqref{eq:B_refiner_residual} for this coefficient yields
\begin{equation}
\begin{split}
  \frac{d\piR_r}{ds_m}={}&\Lambda_m q_r
       -b(Q)q_r\sum_{j\ne r}\frac{dq_j}{ds_m}\\
       &+\left[b(Q)\sum_{j\ne r}\kH_{rj}q_j
                      +(\lambda_r-w)+\zeta_r\right]\frac{dq_r}{ds_m}.
\end{split}
\label{eq:B_dpir}
\end{equation}
For brine, omit both ore terms: $\Lambda_mq_r$ and $(\lambda_r-w)\frac{dq_r}{ds_m}$.

The usual envelope argument eliminates the own-output response from the derivative of a firm's optimized profit at an interior choice. Here, however, I differentiate the refiner's stand-alone operating profit, whereas its output choice reflects ownership interests and sourcing incentives. Its first-order condition therefore does not generally eliminate the own-output term in~\eqref{eq:B_dpir}. For an interior hard-rock plant, $\zeta_r=0$, and the remaining coefficient is $b(Q)\sum_{j\ne r}\kH_{rj}q_j+(\lambda_r-w)$: the first term reflects profits internalized at other plants, and the second is the difference between marginal sourcing cost and the spot ore price. The usual cancellation holds when both terms are zero, or when they offset each other.

\paragraph{Combining the profit responses.}
Weight each producer's profit derivative by its weight in mine $m$'s objective.\footnote{The term \mbox{$\zeta_r\,dq_r/ds_m$} in \mbox{\eqref{eq:B_dpir}} drops out of the decomposition below. Along a smooth downstream equilibrium response with unchanged binding constraints, this product is zero: at an interior output, \mbox{$\zeta_r=0$}; at zero output or fixed capacity, output remains at the same bound, so \mbox{$dq_r/ds_m=0$}.} The resulting marginal payoff separates into a direct term \mbox{$\mathcal D_m$}, a vertical block \mbox{$\mathcal V_m$}, and an upstream horizontal block \mbox{$\mathcal H_m^{\mathrm{up}}$}.

\begin{samepage}
\begin{proposition}[Mine first-order condition]\label{prop:mine_foc}
As the downstream equilibrium varies smoothly with \mbox{$s_m$}, with rival commitments fixed and binding constraints unchanged, the mine's marginal payoff is
\begin{equation}
  \frac{dV_m}{ds_m}=\mathcal D_m+\mathcal V_m
                     +\mathcal H_m^{\mathrm{up}}-\chi_m.
  \label{eq:mine_foc_decomp}
\end{equation}
\end{proposition}
\end{samepage}
The components are defined as follows:
\begin{align}
  \mathcal D_m&=(w-C_m^{M\prime}(x_m))-\Lambda_mx_m,
      \label{eq:mine_foc_direct}\\
  \mathcal R_m^V&=\Lambda_m\sum_{r\in\mathcal R^{\mathrm{HR}}}\kV_{mr}q_r,
      \label{eq:mine_foc_recapture}\\
  \mathcal T_m^V&=-b(Q)\sum_r\kV_{mr}
       \left(q_r\sum_{j\ne r}\frac{dq_j}{ds_m}-\frac{dq_r}{ds_m}\sum_{j\ne r}\kH_{rj}q_j\right),
      \label{eq:mine_foc_tension}\\
  \mathcal S_m^V&=(w-C_m^{M\prime}(x_m))\frac{dA_m}{ds_m}
        +\sum_{r\in\mathcal R^{\mathrm{HR}}}\kV_{mr}(\lambda_r-w)\frac{dq_r}{ds_m},
      \label{eq:mine_source_general}\\
  \mathcal V_m&=\mathcal R_m^V+\mathcal T_m^V+\mathcal S_m^V,
      \label{eq:mine_foc_vblock}\\
  \mathcal H_m^{\mathrm{up}}&=\sum_{k\ne m}\kH_{mk}
                \left[-\Lambda_mx_k+(w-C_k^{M\prime}(x_k))\frac{dA_k}{ds_m}\right].
      \label{eq:mine_foc_hblock}
\end{align}

The decomposition follows by weighting~\eqref{eq:B_dpim}--\eqref{eq:B_dpibr_raw} and substituting~\eqref{eq:B_dpir}. The derivation allows ownership-linked purchases from multiple mines to adjust simultaneously, using the responses $dA_k/ds_m$ implied by the downstream equilibrium.

\paragraph{Interpretation of the components.}
The direct term $\mathcal D_m$ combines the margin on an additional unit of spot sales and the price effect on the mine's total sales.

The vertical block combines the mine's profit change from ownership-linked orders with the downstream profit changes it internalizes:
\[
  \mathcal V_m=(w-C_m^{M\prime}(x_m))\frac{dA_m}{ds_m}+\sum_r\kV_{mr}\frac{d\piR_r}{ds_m}.
\]
Its components separate three effects. \mbox{$\mathcal R_m^V$} captures changes in refiners' ore payments caused by the ore-price response, weighted by the mine's interests in their profits. \mbox{$\mathcal T_m^V$} captures competition effects in the final lithium market as downstream output changes. \mbox{$\mathcal S_m^V$} includes the mine's margin on changes in ownership-linked orders and the refinery-output terms involving the difference between marginal sourcing cost \mbox{$\lambda_r$} and the spot ore price \mbox{$w$}. The vertical block therefore combines payment, sourcing, and competition effects. It cannot be interpreted as a measure of the foreclosure incentive alone.

The upstream horizontal block \mbox{$\mathcal H_m^{\mathrm{up}}$} captures the profit changes at other mines that mine \mbox{$m$} internalizes, arising from changes in ore prices and ownership-linked orders.

\paragraph{Input foreclosure and downstream competition.}
The downstream competition term \mbox{$\mathcal T_m^V$} contains the input-foreclosure incentive:
\begin{equation}
  -b(Q)\sum_r\kV_{mr}q_r\sum_{j\ne r}\frac{dq_j}{ds_m}.
  \label{eq:mine_foreclosure_incentive}
\end{equation}
Holding its own output and sourcing costs fixed, an affiliated refiner loses profit when competing lithium producers expand and lower the lithium price. When additional spot ore induces such an expansion, the mine has an incentive to restrict spot supply to protect the downstream profits it internalizes.

The horizontal weights \mbox{$\kH_{rj}$} in the remaining component of \mbox{$\mathcal T_m^V$} measure how much refiner \mbox{$r$} values producer \mbox{$j$}'s profit. These weights affect how affiliated refiners and their competitors adjust output when more spot ore becomes available. These output responses affect the affiliated refiners' profits, which the mine values through its vertical profit weights. Changing these horizontal weights can therefore alter the mine's incentive to restrict spot supply to protect downstream profits, even when mine ownership and purchasing rights remain unchanged.

\end{document}